\documentclass[jcs,crcready]{iosart1c}     

\usepackage[T1]{fontenc}
\usepackage{times}%
\usepackage[numbers]{natbib}% for bibliography sorting/compressing
\pubyear{0000}
\volume{0}
\firstpage{1}
\lastpage{1}

\usepackage[english]{babel}
\usepackage[fleqn]{amsmath}
\usepackage{amssymb}
\usepackage{euscript}
\usepackage{mathrsfs}
\usepackage{etoolbox} %control structures in latex
\usepackage{float}
\usepackage{subfig}
\usepackage{color}
\usepackage{colortbl} %zum Färben von Spalten und Zeilen in Tabellen
\definecolor{hellgrau}{gray}{0.95}
\usepackage{lscape} %landscape figures
\usepackage{rotating}
\usepackage{multicol} %layout in several columns
\usepackage{epstopdf}
\usepackage{url}

\renewcommand{\emph}[1]{\textit{#1}}

\usepackage{urwchancal}  %Kalligraphische Kleinbuchstabe im Mathemodus: Kommand \mathpzc{}
\DeclareFontFamily{OT1}{pzc}{}
\DeclareFontShape{OT1}{pzc}{m}{it}%
{<-> s * [1.15] pzcmi7t}{}
\DeclareMathAlphabet{\mathpzc}{OT1}{pzc}{m}{it}
\renewcommand{\mathcal}[1]{\mathpzc{#1}}

\usepackage{algorithmic}
\usepackage{algorithm}
\floatname{algorithm}{Algorithm}

\newcounter{casecounter}
\newcounter{subcasecounter}[casecounter] %subcases counter resets on change of cases counter
\newcounter{subsubcasecounter}[subcasecounter]
\newcommand{\formatcaseitem}{}  %Macro zum Setzen des Item Labels in der reasonbycases Umgebung 
\newcommand{\resetcasecounter}{} %Macro zum Zur�cksetzen des Z�hlers in der reasonbycases Umgebung
\newcommand{\selectcasecounter}{} %Macro zur Auswahl des Z�hlers in der reasonbycases Umgebung
\newenvironment{reasonbycases}{

\ifnumcomp{\value{casecounter}}{=}{0}{ %Befehl aus etoolbox Packet
	\renewcommand{\formatcaseitem}{\textit{Case}~\arabic{casecounter}.}
	\renewcommand{\selectcasecounter}{\usecounter{casecounter}}
	\renewcommand{\resetcasecounter}{\setcounter{casecounter}{0}}
}{
	\ifnumcomp{\value{subcasecounter}}{=}{0}{ %Befehl aus etoolbox Packet
		\renewcommand{\formatcaseitem}{\textit{Case}~\arabic{casecounter}--\alph{subcasecounter}.}
		\renewcommand{\selectcasecounter}{\usecounter{subcasecounter}}
		\renewcommand{\resetcasecounter}{\setcounter{subcasecounter}{0}}
	}{
		\ifnumcomp{\value{subsubcasecounter}}{=}{0}{ %Befehl aus etoolbox Packet
			\renewcommand{\formatcaseitem}{\rule[-4.3pt]{4ex}{.52cm}\fbox{\parbox{12ex}{\textbf{Case}~\arabic{casecounter}--\alph{subcasecounter}--\roman{subsubcasecounter}}}}
			\renewcommand{\selectcasecounter}{\usecounter{subsubcasecounter}}
			\renewcommand{\resetcasecounter}{\setcounter{subsubcasecounter}{0}}
		}{
		
		}
	}
}

\begin{list}{
	\formatcaseitem
	}{
	\setlength{\leftmargin}{0ex}
    \setlength{\labelwidth}{0ex}
	\selectcasecounter
	}
 }{
	\resetcasecounter
	\end{list}
}

\usepackage{color}
\usepackage{soul}

\usepackage{listings}

\usepackage{graphicx}
\usepackage{rotating}

\newtheorem{assumption}{Assumption}
\newtheorem{property}{Property}
\newtheorem{example}{Example}
\newtheorem{definition}{Definition}
\newtheorem{proposition}{Proposition}
\newtheorem{theorem}{Theorem}
\newtheorem{lemma}{Lemma}

\newcommand{\system}{\mathcal{Sys}}
\newcommand{\K}{\mathcal{K}}

\newcommand{\pset}[1]{\mathcal{P}(#1)}
\newcommand{\seq}[1]{\langle #1 \rangle}

\newcommand{\pfn}{\hookrightarrow} %partial function
\newcommand{\symbols}{\Lambda}
\newcommand{\symAnd}{ \wedge }
\newcommand{\symOr}{  \vee }
\newcommand{\symNot}{ \neg }
\newcommand{\symBranch}{*}
\newcommand{\symJoin}{+}

\newcommand{\symb}{\mathsf{symb}} %symbolic state
\newcommand{\syminit}{\mathsf{init}} %initialization for symbolic execution

\newcommand{\context}{\mathsf{ctxt}}
\newcommand{\partitions}{\mathsf{tempvs}}
\newcommand{\initpartition}{\mathsf{initv}}

\newcommand{\ftstatus}{\mathsf{st}} % flow Tracker status
\newcommand{\ftidle}{\mathit{idle}} %status idle
\newcommand{\fttracking}{\mathit{tracking}} %status tracking
\newcommand{\ftstop}{\cm{stopFT}} %stop FlowTracker
\newcommand{\ftenv}{\mathsf{sli}}

\newcommand{\hasform}{\ensuremath{\equiv}}

\newcommand{\evalprog}{\mathsf{eval}}  %for the interaction processing program
\newcommand{\evalet}{\mathsf{eval}} %for execution trees
\newcommand{\evalsymb}{\mathsf{eval}} %for symbolic expressions
\newcommand{\trans}{\mathsf{trans}}
\newcommand{\transet}{\mathsf{transToET}}
\newcommand{\transsymb}{\mathsf{transToSym}}

\newcommand{\etlang}{LET}

\newcommand{\partition}{(\block_w)_{w \in \genRange'}}
\newcommand{\block}{\mathcal{B}}

\newcommand{\true}{\mathit{true}}
\newcommand{\false}{\mathit{false}}

\newcommand{\emptytree}{\epsilon}
\newcommand{\nodes}{N}

\newcommand{\Q}[2]{\ifstrempty{#1}{Q}{\ifstrempty{#2}{Q}{#1 \; Q \; #2}}} %correspondence relation

\newcommand{\intprocst}{\mathsf{IP}}

\newcommand{\censorst}{\mathsf{C}}
\newcommand{\flowst}{\mathsf{FT}}

\newcommand{\csep}{\,\mid\,} %separator for components in states

\newcommand{\obsst}{\mathsf{O}}

\newcommand{\vars}{\mathsf{Vars}}

\newcommand{\vals}{\mathit{Vals}}
\newcommand{\high}{\mathsf{high}}
\newcommand{\low}{\mathsf{low}}
\newcommand{\hvars}{\mathsf{Hvars}}
\newcommand{\lvars}{\mathsf{Lvars}}
\newcommand{\lequiv}[2]{\ifstrempty{#1}{\obsst}{\ifstrempty{#2}{\obsst}{\obsst #1 =  \obsst #2}}}   %\sim

\newcommand{\bstate}{\mathcal{ais}}
\newcommand{\bstates}{\mathcal{AIS}}

\newcommand{\basicreq}{\mathsf{br}}
\newcommand{\basicreqs}{\Pi}
\newcommand{\op}{\oplus}
\newcommand{\eval}{\mathsf{eval}} %evaluation of operation

\newcommand{\conf}{\mathsf{pol}}

\newcommand{\mem}{\mathsf{mem}}

\newcommand{\view}{\mathsf{prev}}
\newcommand{\infintbel
}{\mathcal{IAI}}
\newcommand{\DS}{\ensuremath{D\! S}}
\newcommand{\ID}{\ensuremath{I\! D}}

\newcommand{\env}{\vdash_\Gamma}
\newcommand{\secret}{\mathcal{S}}

\newcommand{\cm}[1]{\mathsf{#1}}  %formating of commands
\newcommand{\code}{\cm{p}}
\newcommand{\intprocprog}{\cm{p}}
\newcommand{\intproclang}{\ensuremath{L\! I\! P}} %language for interprocessing programs
\newcommand{\emptyprog}{\epsilon}

\newcommand{\censor }{\cm{censor}}

\newcommand{\ifc}[3]{\cm{if}\;#1 \; \cm{then}\;#2 \; \cm{else}\; #3 \cm{endif}}

\newcommand{\icontrol}[2]{\ifstrempty{#1}{\cm{decl}}{\ifstrempty{#2}{\cm{decl}}{\ensuremath{\cm{decl}(#1,#2)}}}}
\newcommand{\disttab}{\mathsf{dt}} %distortion table
\newcommand{\genRange}{R} %range of values for generalization
\newcommand{\secconf}{S\! C} %security configuration
\newcommand{\vioSets}{V} %disclosure

\newcommand{\topg}{ANY}
\newcommand{\taxtree}{T}
\newcommand{\gscheme}{\ensuremath{\mathcal{G}}}
\newcommand{\gtree}{G}
\newcommand{\gle}{\le}

\begin{document}
%
%\frontmatter          % for the preliminaries
%

\pagestyle{headings}  % switches on printing of running heads

\begin{frontmatter}  

\title{Confidentiality enforcement\\by hybrid control of information flows}

\runningtitle{Confidentiality enforcement by hybrid control of information flows}  

\author{\fnms{Joachim} \snm{Biskup}},
\author{\fnms{Cornelia} \snm{Tadros}} and 
\author{\fnms{Jaouad} \snm{Zarouali}}
\runningauthor{J. Biskup et al.}
\address{Fakult\"at f\"ur Informatik, Technische Universit\"at Dortmund, Germany\\
 E-mail: \{joachim.biskup|cornelia.tadros|jaouad.zarouali\}@cs.tu-dortmund.de}

\maketitle                % typeset the title of the contribution

\begin{abstract}
 \hspace*{1ex}An information owner, possessing diverse data sources, might
want to offer information services based on these sources to cooperation partners 
and to this end interact with these partners by receiving and sending messages,
which the owner on his part generates by program execution.
Independently from data representation or its physical storage,
information release to a partner might be restricted by the owner's confidentiality policy
on an integrated, unified view of the sources.
Such a policy should even be enforced if the partner as an intelligent and only semi-honest attacker attempts to infer 
hidden information from message data, also employing background knowledge.
For this problem of inference control, we present a framework for a unified, holistic control of information flow induced by program-based processing of the data sources to messages sent to a cooperation partner.
Our framework expands on and combines established concepts for confidentiality enforcement and its verification and is instantiated in a Java environment. 
More specifically, as a hybrid control we combine gradual release of information via declassification, enforced by static program analysis using a security type system, with a dynamic monitoring approach.
The dynamic monitoring employs flow tracking for generalizing values to be declassified under confidentiality policy compliance. 
\end{abstract}   


\end{frontmatter}  

\section{Introduction}

Today's IT-security technologies provide a broad variety of effective and efficient mechanisms 
to prohibit unauthorized reading of any kind of raw \textit{data}, 
e.g., authenticated access control and private-key or certified public-key encryption. 
And these technologies also offer somehow limited approaches
to confine the \textit{information} content of data made accessible to cooperation partners, e.g., 
language-based information flow control, information systems with inference-usability confinement, 
confidentiality-preserving data publishing and cryptographic multiparty computations.
However, independently of its carrier and its representation, information is the fundamental \textit{asset} 
of an individual pursuing self-determination or an enterprise
doing business. 
Moreover, information arises not only from accessible data 
but essentially also from a priori \textit{knowledge} and \textit{intelligence} of a
reasoning observer and, additionally, is \textit{accumulated} over the time.
So, being widely unsolved so far, the challenge is 
to enable an individual or an enterprise as an information \textit{owner} to exercise a \textit{holistic control} over the information conveyed to communication partners by means of transferred data.

We address this challenge in the following narrower, yet comprehensive scenario. 
The owner's information basis is modeled as an \textit{abstract information state} which might be a single relational database, a collection of diverse data sources
or a dedicated virtual view derived from a federation of heterogeneous data sources.
This basis is processed program-based by a single control component, called \textit{mediator}, to determine reactions to requests received from a cooperation partner 
of the information owner under  selective prohibitions of information release stated in the owner's \textit{confidentiality policy}.
The mediator's approach is twofold, first, to explore the options of the partner, seen as a semi-honest attacker, to exploit the communicated data to infer information to be kept hidden, 
and, second, to block all such options by filtering and modifying the data appropriately. 

Thereby, the exploration of the attacker's inference options should be based on his (assumed) general background and a priori knowledge as well as on the tracing of the overall history and  the tracking of  the current control flow of the mediator's program execution. 
In particular,  we focus on the partner being an initiator of requests, an observer of reactions and a rational attacker against the policy, but do not consider his client system by means of which he interacts with the mediator.

The policy only specifies \textit{what} should be kept confidential
and leaves open %to the monitoring mechanism 
\textit{how} confidentiality will be enforced.         
For an  enforcement %as shown in Figure~\ref{fig:infcontroloutline}, 
we propose a two-layered approach of employing 
existing concepts, namely 
\begin{itemize}
\item language-based \textit{information flow control} with declassification for the program-based interaction processing, e.g.~\cite{SaSa09}, and
\item \textit{inference-usability confinement} with dynamic tracking of information flows from the abstract information state to generated message data sent to the partner and with a filter and modification method that employs generalization of declassified values, e.g.~\cite{Bi12a,Bi13,BiBoGaSa14}. 
\end{itemize}
Our main contributions are
\begin{enumerate}
\item a fundamental framework for unified, holistic control of information flow through program execution from an abstract information state to the partner, 
\item a formal verification that the control enforces the confidentiality policy under restrictions on the expressiveness of the programming constructs involved, and
\item a Java-based, exemplary instantiation of that framework.
\end{enumerate}

%Our verification method 
The proposed control bases on the \textit{gradual release} property~\cite{AsSa07} established by security type systems for language-based information flow control. 
According to this property,  program execution may release information from the abstract information state, but exclusively and gradually via dedicated declassification assignments, and thus via no other program constructs. 
The guarantee of this gradual release property by  a suitable security type system is the basis of our verification of confidentiality enforcement.
Additionally, what information is declassified is controlled by a dynamic monitor for confidentiality-policy compliance in the spirit of inference-usability confinement.

Inference-usability confinement does not confine information disclosure, but such that would enable the partner to infer a confidential piece of information.
The key method to do so is to simulate the partner's options for inference by two means,
knowledge updates and entailment tests.
First,  a control mechanism for inference-usability confinement may update a usually logic-based model of the partner's knowledge
according to what knowledge the partner may gain through a specific message sent to him as a reaction. 
Second, the control mechanism may test whether such an updated knowledge model entails a confidential piece of information
(in the pertinent logic).
Employing these means, including possibly several entailment tests each based on a specific, tentative knowledge update, the control mechanism may decide for or against confining the information content of the message to be sent to the partner.
 
In our two-layered approach, the dynamic monitor simulates the partner's inference similarly, because it also employs a kind of knowledge update and a kind of entailment test. 
First, the monitor tracks information flow for determining what knowledge the partner may gain through a declassified value and for updating 
a model of the partner's knowledge accordingly.
Second, the monitor tests whether the disclosure of pieces of tracked information gives the partner the option of harmful inference.
By these means, the dynamic monitor decides whether to filter information (as a confinement) by modifying a declassified value to a suitably generalized value. This way, the monitor extends the control by the type system for gradual release, by determining what information is to be declassified, and confining 
this information to that conveyed by a suitably generalized value if necessary
for confidentiality policy compliance.

This combined control by both the type system and the monitor is shown to be effective by adopting a proof method from~\cite{BaNaRo08}, a modular static approach for conditioned gradual release. 
The conditioned gradual release property extends gradual release with a declaration of what information may be declassified (in relational logic).
A program may be verified to have this property by a static analysis of its code. 

As part of the framework, we also provide a general scheme by means of which declassified values are generalized to confine the information disclosed. And furthermore, inspired by symbolic execution~\cite{BaDaGu12}, we present an algorithmic approach for the monitor's dynamic tracking of abstract information.
Our contributed framework leaves room for many optimizations, for example an apt balance between static program analysis and dynamic monitoring. Such optimizations can be explored reasonably given a concrete class of abstract information states 
and could be built on results of the established methods we employ for our framework.
In the conclusion of this article,  setting our contribution into the context of related work, we outline ideas how to do so.

\noindent\textbf{Outline of article}.
	
In Section~\ref{Sec:Overview},  we overview the essential design concepts of the mediator framework  
according to our two-layered approach and illustrate them with a running example (part of the first main contribution).

In Section~\ref{Sub:Impl}, we instantiate the framework in a Java environment, exemplifying the
challenges inherent in any instantiation and suggesting respective preliminary solutions (part of the third main contribution).
Moreover, this instantiation demonstrates the joint functioning of the framework's components introduced in Section~\ref{Sec:Overview}.

In Section~\ref{Subsec:Mediator}, we define the functionality of the desired mediator as part of a run-based system  under the objective of a unified, verifiable declaration of the whole framework.
The other part of this system models an observer, which as a worst case assumes the capabilities of the cooperation partner as an attacker
against the confidentiality policy. 
This model of the observer is presented in Section~\ref{Sec:Observer} and completes the declaration of the semantic property for confidentiality,
outlined in Section~\ref{Sec:Overview}. 
Afterwards, in Section~\ref{Sec:Declarative} we detail the control by the dynamic monitor based on declarative properties that we motivate and introduce in the same section for the tracking of information flow from the abstract information state to program data. 
Then,  we verify that the declaratively specified mediator enforces the confidentiality property  (first and second main contribution).
The proofs for this verification are shifted to the appendix. 

As a further elaboration of the mediator framework (first main contribution), in Section~\ref{Sec:DistTab},
we present an algorithm for determining a generalized value based on a summary of the observer's options of inference as prepared by the dynamic monitor. This algorithm implements a declarative concept from Section~\ref{Sec:Overview} as we prove,
and additionally relies on a generalization hierarchy of values similarly as anonymization algorithms for \textit{k}-anonymity. 
Furthermore, in Section~\ref{Sec:FlowTracker}, adapting ideas from symbolic program execution, we present algorithms that implement the declarative requirements for the tracking of information flow from Section~\ref{Sec:Declarative} and verify that these requirements are indeed achieved.

In Section~\ref{Sec:FurtherChallenges}, we present selected challenges for instantiating components of the framework introduced 
from Section~\ref{Subsec:Mediator} on and sketch respective solutions of our Java-based instantiation introduced in Section~\ref{Sub:Impl}.
Finally, in Section~\ref{Sec:RelatedWork},  by additionally considering related work, we will discuss our contributions.

\section{Overview on the framework}\label{Sec:Overview}

\begin{figure}
\begin{center}
\def\svgwidth{\textwidth}
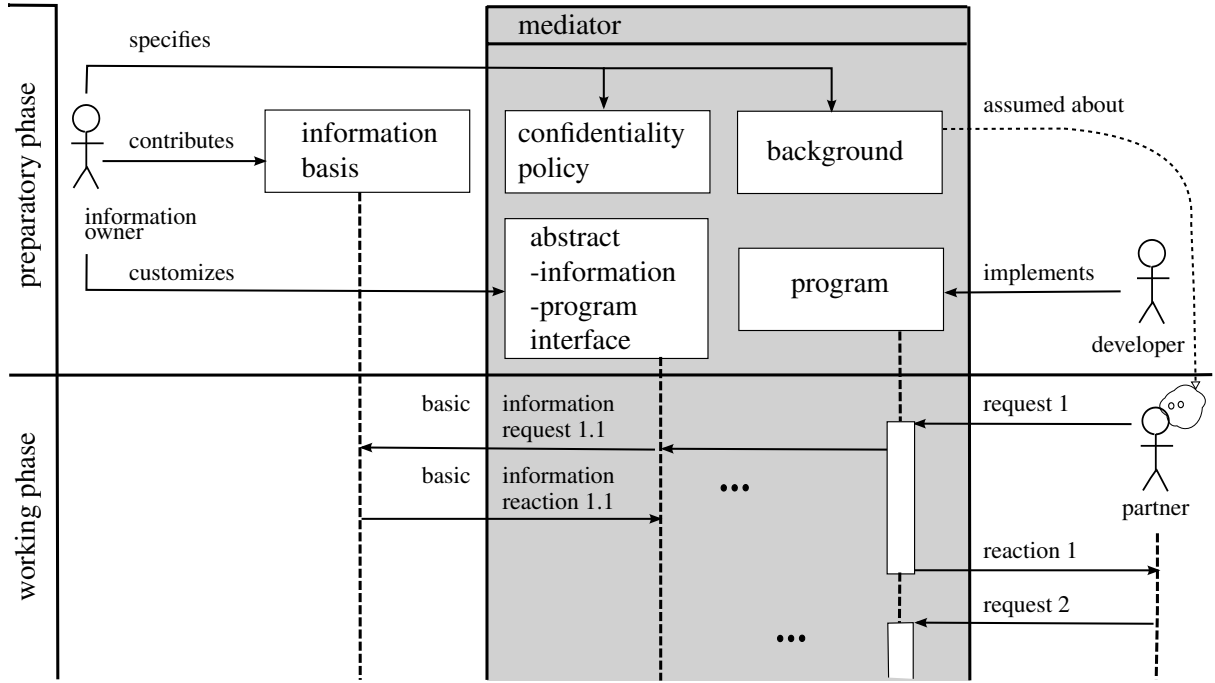
\vspace*{1.6cm}
\end{center}
\caption{Scenario of an information owner cooperating via a mediator with one partner 
            with the owner's concern for the confidentiality of information contained in his information basis}\label{Fig:Scenario}
\end{figure}

The main goal of the framework is confidentiality preservation for the mediated abstract information state processed program-based as an information service to the partner by \textit{uniformly} and \textit{securely relating} two established methods of control.
On the one hand, the first method, inference control in the spirit of inference-usability confinement, controls the information release from the abstract information state by dynamic tracking and value generalization.
On the other hand, the second method, information flow control,  considers information with different degrees of sensitivity and contained in data objects which are stored in variables and are manipulated by means of commands like evaluation of expressions, assignment or conditionals (guarded commands). 
In this section, first we will make more precise the information mediation scenario for which we aim to achieve this goal
as shown in Figure~\ref{Fig:Scenario}. 
Then, we will introduce our proposed framework step-wise along the line of reasoning that justifies our approach. 
In the next section, we will summarize the introduced components of the framework by their comprehensive formalization as a run-based system and give an example of instantiating the framework. 

\noindent\textbf{Scenario}. 
Preparing the mediator's information services, a developer, on behalf of the information owner, implements a \textit{program} which generates reactions to requests by a specific cooperation partner,
for example, to make sales offers according to requests by the partner as a buyer.  
For these services, the information owner has defined a dedicated \textit{abstract-information-program interface}
via which the program may query the abstract information state, the \textit{information basis} contributed by the information owner. 
Such queries, here called \textit{basic information requests},  are the only way how the program may get data from the abstract information state. 
Moreover, the information owner declares a partner-specific \textit{confidentiality policy} $\conf$ as sets of abstract information states, as described in~\cite{BiBoGaSa14}. 

\begin{example}\label{Exa:Scenario}
As a simple running example we consider an abstract information state in form of a relation $\DS$  over the attribute $\ID$ as a primary key and further functionally dependent attributes $A,B$ and $C$. 
Each attribute $X$ has a finite domain $\mathit{dom}(X)$ assumed to be disjoint with the domains of the other attributes.
Via the interface the mediator's program can query the state by unnested select or project queries, where selection predicates are restricted to conjunctions of attribute-equals-value terms. 
To keep it simple, the program only accesses the state for a single row as identified by its $\ID$ value $id$.
We denote this row by $\DS^{id}$.
The value $id$ can be thought of as an identifier of an individual and the row as data related to him.
With regard to this individual's privacy, the data owner  specifies as confidential\\
 that $C$ has the value $c_1$ by $\secret_1 = \{(id,\mathit{A},\mathit{B},c_1) \mid \mathit{A}\in \mathit{dom}(A), \mathit{B}\in\mathit{dom}(B) \}$,\\ that $C$ has the value $c_2$  by  $\secret_2 = \{(id,\mathit{A},\mathit{B},c_2) \mid \mathit{A}\in \mathit{dom}(A), \mathit{B}\in\mathit{dom}(B) \}$\\ and that the value $b_2$ of $B$ occurs with the value $c_3$ of $C$ by  $\secret_3 = \{(id,\mathit{A},b_2,c_3) \mid \mathit{A}\in \mathit{dom}(A) \}$.
\end{example}

As a threat to this policy, 
the partner is seen as a semi-honest attacker who always uses requests with a valid format and type, but who may employ
a sequence of such requests to find out confidential information.
The information owner specifies the \textit{background} which the partner might exploit for this purpose,
assuming the partner's actual a priori knowledge.

During the mediator's service, the program is executed by the mediator to handle the partner's requests on the one hand, and to keep selected pieces of such information confidential from the partner as declared in the policy on the other hand. 
More specifically, being rational, the cooperation partner might reason about the reaction data transmitted to him by a single message generated by one execution of the program based on the actual abstract information state and the partner's request.
%, as caused by the mediator's interaction processing by the pertinent program execution on the actual abstract information state, and by 
By reasoning thus the partner might infer that this state contains a certain piece of confidential information. 
Moreover, for such a reasoning the partner might also inspect the whole \emph{history} of message data generated by possibly several subsequent program executions.
In this way, the partner might figure out that the set of states appearing possible to him (which contains the actual state)
is a subset of at least one of the states $\secret$ specified as an element of the confidentiality policy 
(and thus the actual state is contained in such a sensitive $\secret$).

The partner's capabilities as an attacker will be assumed as those of a potentially more powerful
\emph{observer} which may not only perceive message data as the partner may do, but in a limited way may perceive the mediator's progress of computation as the partner might not do.
The capabilities of this observer will be formalized as part of a system consisting of the mediator and the observer.
This formalization abstracts the mediator's activities and the observer's perceptions thereof to a function, called run, from time, represented as $\mathbb{N}_0$, to the mediator's and the observer's states. 
The (formal model of the) system $\system$ then is a set of runs and has several parameters among which are the mediator's program  $\intprocprog$, an abstract information state as an element of a set $\bstates$ of abstract information states,  a confidentiality policy $\conf \subseteq \pset{\bstates}$,  and program input from a single request by the partner.

Based on this formalization, the observer's inference will be modeled by a knowledge operator $\K$ about possible abstract information states. 
Finally, the confidentiality policy requires that the observer, and thus the less powerful partner, should not be able to infer
a confidential piece $\secret$ of information.
Thus, we aim at the following property.
\begin{property}[Confidentiality preservation]\label{Def:Conf}
For all runs $r \in \system$ and times $t$ and for all $\secret \in \conf$ it holds
$ \K (r,t) \not \subseteq \secret $.
\end{property}

\begin{figure}[t!]
\begin{center}
\includegraphics[width=  1.0 \linewidth]{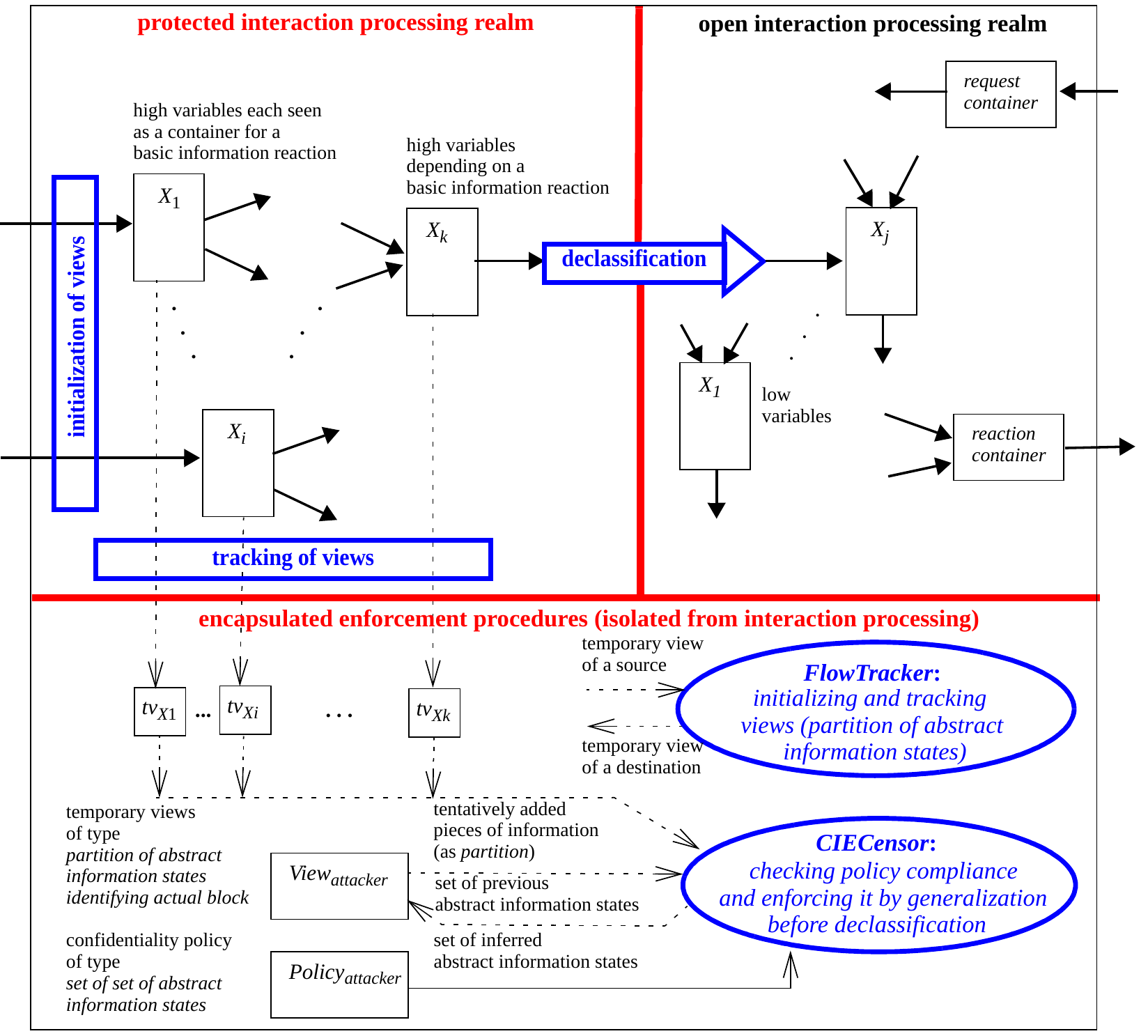}
\end{center}
\caption{Components of the proposed framework with a two layered approach according to the draft in~\cite{BiTa15}: Release of information from the abstract information state to the partner solely through declassification, and control of the information flow through declassification for confidentiality policy compliance by a dynamic monitor which consists of FlowTracker and CIECensor}\label{Fig:Concept}
\end{figure}

The sole purpose of the mediator framework is to enforce this property in an automated manner,
filtering and modifying as few pieces of information as possible for a most informative reaction to the partner.
For this purpose,  the mediator employs a \emph{flow tracker} for dynamic tracking of information flow from the abstract information state
to data processed by the program and a \emph{censor} for policy-compliant value generalization.
The jointly coordinated functioning of the flow tracker and the censor as a dynamic monitor of declassification assignments is illustrated by Figure~\ref{Fig:Concept} and explained in the remainder of the section.
The remainder of the section is further divided into items along the line of reasoning supporting our approach.

\begin{example}\label{Exa:Prog}
For the upcoming exposition of the mediator framework, consider the following purposeless program $\intprocprog$
the arguments $arg_1, arg_2$ of which are set via the partner's request and the 
return value $x_{rea}$ of which is forwarded to the partner.
A basic information request $\basicreq$ to the interface takes the query kind and
the respective parameters as arguments.
The marked lines are being explained in the sequel.
\begin{small} 
\begin{gather*}
\intprocprog (arg_1,arg_2): x_{rea}\\
\text{1: } \ifc{arg_1 \; \cm{IN} \; \mathit{dom}(C)}{\underline{x_1 := \cm{NOT} \; \cm{ISEMPTY} (\basicreq (\text{select} , C=arg_1))}\\\text{2: } }{x_1:= \cm{FALSE}\;} \\
\text{3: } \ifc{arg_2 \; \cm{IN} \; \mathit{dom}(C)}{\underline{x_2 := \cm{NOT} \; \cm{ISEMPTY} (\basicreq (\text{select} , C=arg_2))}\\\text{4: } }{x_2:= \cm{FALSE}\;}\\
\text{5: } \underline{x_3 :=  \basicreq (\text{project}, \{A,B\})} \\
\text{6: } \underline{x_4 :=  \basicreq (\text{project} , \{A,C\})} \\
%\begin{aligned}
\text{7: } \underline{
\ifc{ x_1 \; \mathsf{or} \; x_2}{ 
x_5:= x_3
}{x_5:=x_4\;}}\\
\text{8: } \icontrol{ x_5 }{x_6}\\ 
\text{9: } x_{rea}:=\cm{TOSTRING}(x_6)
%\end{aligned}
\end{gather*}
\end{small}  
\end{example}

\noindent\textbf{1. Isolation by typing}. 
The working space of the mediator's program has to be strictly divided into 
a \textit{protected realm} where potentially confidential information stemming from reactions to basic information requests, basic information reactions for short, is processed 
and an \textit{open realm} where the final external reaction to the partner's request is prepared. 
The division into realms is virtually achieved by separating the set $\vars$ of program variables into two subsets, that of low variables $\lvars$ to which only data processed in the open realm may be written as a preliminary rule, and that of high variables $\hvars$, without any write restrictions.  

Technically, the mediator's program will be typed by means of a security level inference $\env$ when the developer compiles the program.
This level inference starts with the basic set $\Gamma$, which assigns levels from $\{\high , \low \}$ to selected expressions, and inductively derives levels of further expressions, including program variables,  and levels of assignments as elementary programs and of their composition to complex subprograms by control flow constructs. 
Our approach is not to select a specific security level inference system\footnote{
Or more generally, we do not rely on  a  specific security type (inference) system,  which may assign and derive more expressive types than security levels in the mentioned inductive way.  But we do not need the enriched expressiveness for our exposition.
}, but to point to properties needed for the verification of our framework and offered by numerous such systems to-date.  
Each time we eventually instantiate the framework, we select the security level inference system as we demonstrate in Section~\ref{Sub:Impl}.
  
The meaning of variable levels, as we desire for the division of processing realms, are phrased more concisely by two rules,  adhered to by level inference. The first rule \emph{no-read-up} roughly says that a subprogram typed low, denoted by $\code: \low$, may not read high-level variables. Complementarily, the second rule \emph{no-write-down} roughly says  that a subprogram typed high, denoted by $\code:\high$, may not write to low variables. As stated in the strict form here, these rules applied together ensure that the value of a high variable does not affect the values of low variables at all. 

This lack of effect of high processing on low processing can be stated as a security property like non-interference~\cite{SaSa09}.
If a security level inference guarantees such a property, then it actually achieves the desired isolation of realms under three further restrictions.
First,  basic information reactions must be treated as high variables by level inference. 
Second,  a designated variable  as a container for the reaction sent to the partner must be low.
Third,  to largely simplify the design of the dynamic monitor we further require that parameters to basic information requests must be low.
In formal terms, these requirements could be specified in the set $\Gamma$ on which level inference $\env$ bases. 

\begin{example}
In the program code displayed in Example~\ref{Exa:Prog} parts processed in the protected realm
are underlined: these are those parts that may not write to low variables. 
The variables $x_1$ to $x_4$ get assigned the return values of basic information requests and therefore
may not be low. 
The guard of the if-statement in line~7 thus evaluates high variables.
Since this evaluation may not affect the change in low variables, they may not be written in the scope of the guard.
\end{example}

\noindent\textbf{2. Sharing by declassification}.
To nevertheless enable \textit{discretionary sharing} of information about the abstract information state, i.e., 
	a controlled information flow from a variable in the protected realm to a variable in the open realm, 
	we will use \textit{declassification}  by means of an explicit assignment  $\icontrol{x_{src}}{x_{dest}}$ for declassification of the content of variable $x_{src}$ to variable $x_{dest}$, to be offered by the level inference system used.
This is the sole means to let information contained in the abstract information state flow from the protected to the open realm.

Absence of (exploitable) information flow about this state via other program constructs than a declassification assignment means that after the execution of such a program construct, but declassification, the partner cannot extend his knowledge.
This restriction on the way how the partner may extend his knowledge about the abstract information state
can be enforced by the security level inference system as the property of gradual release.
Hence, we aim at the following property that considers the  knowledge before and after the one-step execution of the next program construct, called \textit{active command}. 
\begin{property}[Gradual release, cf.~\cite{AsSa07}]\label{Def:Gradual}
For all runs $r \in \system $ and times $t$ where the active command in $(r,t)$ is not a declassification assignment,
 it holds $\K (r,t+1) = \K (r,t).$
\end{property}

\begin{example}
In the program code of Example~\ref{Exa:Prog} a value contained in a high variable 
can only be assigned to the variable $x_{rea}$ for the reaction to the partner if this value
has been declassified before. 
A rule of thumb is that the earlier in the code a value is declassified the less informative a possible
modification of that value is as a negative effect, and the less computational effort the
dynamic monitor requires as a positive effect.  
This rule will be illustrated in Section~\ref{Sec:FurtherChallenges}.
\end{example}

\noindent\textbf{3. History-aware policy compliance by FlowTracker and CIECensor}.
	\textit{Before} transferring data through the declassification assignment 
  the mediator has to ensure that the transfer would be harmless, i.e., complying
	with the confidentiality policy under the simulated history-dependent \textit{previous view} $\view \subseteq \bstates$ of the attacker on the mediator's abstract information state. 	For this complex task, we will provide a dedicated encapsulated component called \textit{FlowTracker}
	which can delegate subtasks, namely (1) evaluations of harmlessness and (2) actions of filtering and modifying,
	to  a further component called \textit{CIECensor}.
	
	The previous view $\view$ describes knowledge of the partner about the abstract information state in the following sense: The partner can rule out any state in $\bstates \setminus \view$ as the actual abstract information state $\bstate$ without being mistaken since $\bstate \in \view$ will be an invariant throughout interaction processing guaranteed by the proposed framework.
Initially, this view can be taken from the previous execution of the interaction processing program for the cooperation partner or, if there is no such execution,  initialized to a definition of a priori knowledge for that partner.
To simplify notation,  we define $\bstates$ accordingly and let the mediator initialize $\view := \bstates$.
Intuitively, initial policy compliance should amount to ensuring $\view \not \subseteq \secret$  for all $\secret \in \conf$.
With the previous simplification, this requirement can be reformulated to $\secret \neq \bstates$.

\begin{example}\label{Exa:AIS}
Since the partner knows, in our simplified illustrative scenario, that the mediator accesses the row of an individual
identifiable by its $\ID$ value $id$, the set $\bstates$ is the set of all such rows $\{(id,\mathit{A},\mathit{B},\mathit{C}) \mid \mathit{A} \in \mathit{dom}(A), \mathit{B} \in \mathit{dom}(B), \mathit{C} \in \mathit{dom}(C)   \}$.
Obviously the policy given in Example~\ref{Exa:Scenario} is not compromised by this initial knowledge.
\end{example}
 
\noindent\textbf{4. Need of local flow tracking}.
For an \textit{evaluation of harmlessness} of a requested data transfer through a declassification assignment, the CIECensor needs  
	to know both the explicit and the implicit information about the abstract information state contained
	in the source variable $x_{src}$ as the result of the preceding processing in the protected realm.\footnote{
	  As it is common~\cite{BaNaRo08,BrDeSa13}, we assume that $x_{src}$ is not declassified in the scope of a high level guard so that only the preceding evaluation of the guard of a guarded command, the execution of which has been completed,  may be conveyed implicitly through the declassification.
	  The only way how this implicit information may be conveyed is by the value of $x_{src}$.
	}
	 I.e., the CIECensor
	needs an additional input, namely a temporary view on the abstract information state for $x_{src}$, solely resulting from preceding processing
	and to be tentatively combined with the attacker's previous view.     
	This \textit{tentative addition} will be dynamically generated by the FlowTracker.\\[1ex]

\noindent\textbf{5. Identifying implicit flows by symbolic executions}.
	%Since i
	Implicit information -- as caused by guarded commands -- not only depends on the  execution path actually performed
	but also on the (alternative) paths possibly followed which could be selected for different values of the pertinent guards.
	Accordingly, the FlowTracker will base the dynamic generation of the tentative addition, the temporary view on the abstract information of a variable $x_{src}$ to be declassified, on a  
	\textit{symbolic expression}. 
	This symbolic expression stems from \textit{symbolic program execution} of a piece of code, which computes the value of $x_{src}$
	and has a high security level, 	as inspired by~\cite{BaDaGu12}. 
	
	Symbolic program execution will translate the considered piece of code to an execution tree, the leaves of which each represents a set of possible execution paths.
	Throughout \textit{all} paths of this tree, symbolic program execution will follow the changes of high variables and denote 
	their respective information contents, regarding the abstract information state, by symbolic expressions.
	These expressions may refer to each program expression,  the initial value of which is used in a path, by a respective (still uninterpreted) symbol. 
	
	\begin{example}\label{Exa:SymbExpr}
	The if-statement in line~7 of the code of Example~\ref{Exa:Prog}, when considered in isolation,  is translated to the symbolic expression  	$(x_1 \symOr x_2) \symBranch x_3  \symJoin \symNot (x_1 \symOr x_2) \symBranch x_4$ for the information content of variable $x_5$ regarding the abstract information state.
	The symbol $\symJoin$ separates the alternative paths, while the symbol $\symBranch$ connects boolean conditions/computations along each path. The symbols $\symNot, \symOr$ denote boolean operators. The interpretation of all these symbols is illustrated in the next step.
	\end{example}
	
\noindent\textbf{6. Determining local flows by FlowTracker}.
	To prepare for declassification, 
	%to enable declassification, 
	the \textit{FlowTracker} will evaluate the symbolic expressions
	 for high variables manipulated by a considered piece of code
	using the actual information contents of other variables involved.
	%whose denominations appear 
	%that are referred to in the annotation. 
	These information contents
	are dynamically determined as \textit{temporary views} associated with these variables. 
	More specifically, these temporary views are 
	\textit{initialized} when storing a basic information reaction in a variable seen as a container and 
	for an evaluation further \textit{tracked} according to the operators in the symbolic expressions. 
	
	In abstract terms, such an information content is represented as an $\genRange$-indexed partition of $\bstates$,  similarly as in~\cite{BiBoGaSa14}, which is a partial function of type $\genRange \pfn \pset{\bstates}$ such that the elements of its image form a partition of $\bstates$ and $\genRange$ is a finite subset of values $\vals$. We write $(\block_w)_{w\in \genRange'}$ where  $\block_w \subseteq \bstates$ is the image of $w$ and $\genRange' \subseteq \genRange$ is the domain of the function.
A temporary view containing $(\block_w)_{w\in \genRange'}$ for variable $x$ means the following:
knowing
the value of  $x$ is $w$ enables the attacker to infer to which block the actual abstract information state belongs, namely $\block_w$.	
	Moreover, the partition is complemented with an identification of the \textit{actual value} of the variable and \textit{block} of the partition, respectively.

\begin{example}\label{Exa:TempView}
Reconsider the program of Example~\ref{Exa:Prog} called with values $c_1$ and $c_3$ obtained from the partner's request.
In line~7,  variable $x_5$ contains the result of projecting the abstract information state to attributes $A$ and $B$ in the if-branch, or $A$ and $C$ in the else-branch.
The step-wise construction  of the temporary view for $x_5$ for this program line is illustrated by Figure~\ref{Fig:Partition} and explained in
the following.

If we take $\mathit{dom}(C) = \{c_1,c_2,c_3,c_4\}$, a projection to attributes $(A,C)$ partitions $\bstates$ of Example~\ref{Exa:AIS} into the blocks above 
the dotted crossline. 
These blocks are represented by four boxes each labeled with a set of indexes. 
Each index, which is of the form  $(\mathit{A},\mathit{C}), \mathit{A}\in \mathit{dom}(A), \mathit{C} \in \{c_1,c_2,c_3,c_4\}$, refers 
to a respective block, in total yielding $4 \cdot \mathit{dom}(A)$ many blocks above the crossline.
Likewise, a projection to $(A,B)$ yields a partition as below the dotted crossline where the single box represents  $\mathit{dom}(A) \cdot \mathit{dom}(B)$ many blocks.
\begin{figure}[h]
\begin{center}
\includegraphics[width=\textwidth]{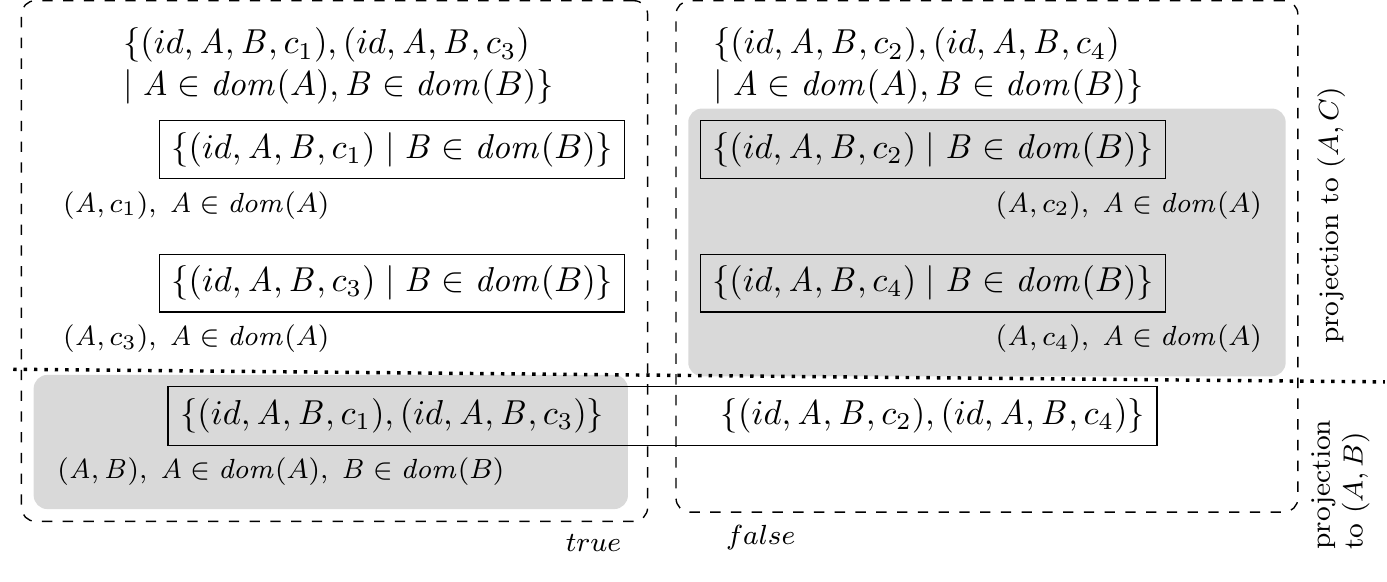}
\vspace*{-1cm}
\end{center}
\caption{Partitions for the flow tracking
through the if-statement in line~7
to variable $x_5$:
the initial partition for the guard 
$x_1 \; \mathsf{or} \; x_2$ consists of the true/false-indexed blocks in dashed lines dividing
$\bstates$ into the two sets in the first line of the diagram}\label{Fig:Partition}
\end{figure}

If the cooperation partner observed the so computed value of $x_5$, he could single out the actual abstract information state as a member of a specific block in the grayly shaded area where blocks below the dotted crossline are reduced to their intersection with this area.  For example, if that value was the pair $(a_1,b_2)$, the partner may conclude that the state is one of $\{(id,a_1,b_2,c_1),(id,a_1,b_2,c_3)\}$.

Now, let us see how the FlowTracker determines these blocks in the grayly shaded area by interpreting the symbolic expression $(x_1 \symOr x_2) \symBranch x_3  \symJoin \symNot (x_1 \symOr x_2) \symBranch x_4$ introduced in Example~\ref{Exa:SymbExpr}.
As for the subterm $(x_1 \symOr x_2) \symBranch x_3$, the FlowTracker evaluates symbol $x_3$ to the temporary view of $x_3$ which contains the partition below the dotted crossline. 
Then, the symbol $\symBranch$, which connects this evaluation with the path condition $x_1 \symOr x_2$, is interpreted as intersecting the obtained partition with the true-indexed block drawn with dashed lines. 
The previous evaluation of the symbol $\symOr$ in the path condition 
is according to a comprehensive definition applicable to any n-ary operator, boolean and non boolean,  
and discussed in Section~\ref{Sec:FlowTracker}.

Likewise, the FlowTracker evaluates the other subterm.
Finally, evaluating the symbol $\symJoin$, the FlowTracker combines the two partitions, one for each of the considered subterms, by a union of the two respective sets of blocks, which are those in the grayly shaded area.
The set union corresponds to the partner's considering each of the two alternative paths in line~7 as a potential candidate for the actual execution path. If the partner observes the actual value of $x_5$ and then uses the partition in the grayly shaded area,  he may figure out the actual path due to the domains of $B$ and $C$ being disjoint, but this is not necessarily so.
\end{example} 	

\noindent\textbf{7. Evaluation of harmlessness by CIECensor}.
  Provided by the \textit{FlowTracker} with the tentative addition,  for declassification the \textit{CIECensor}
	checks whether the \textit{combined information content} of the tentative addition resulting from preceding processing and the previous view resulting from the history 
	could possibly violate the \textit{confidentiality policy}.  
	In abstract terms, this combination 
	%is defined to be \textit{greatest lower bound} of the partitions involved
	is obtained by taking all nonempty intersections of a block in (the partition of) the tentative addition with
	%(the single block of) 
	the previous view, 
	%of a block in one partition with a block in the other partition), 
	and a \textit{possible violation} occurs
	if there is a block in the combination that is \textit{completely contained} in an element of the confidentiality policy.
	If the check confirms harmlessness, the previous view is updated to its intersection with the actual block and the declassification processed without filtering and modifying.
	
	Otherwise, such a harmful situation can be described by a non-empty \textit{security configuration} which the CIECensor will use for generalization of the value to be declassified.
\begin{definition}[Security configuration, cf.~\cite{BiWe08}]\label{Def:SecConfig}
Let $\partition$ be an $\genRange$-indexed partition of  $\bstates$, 
let $\view \subseteq \bstates$ be the previous view and 
let  $\conf \subseteq \pset{\bstates}$ be a confidentiality policy.
Then, the security configuration $\secconf$  consists of the domain $\genRange'$ and a set $\vioSets$ (violating sets) of sets  $ I \subset \genRange'$ of block indices defined by the condition
\begin{equation*}
 \text{exists }\secret \in \conf \text{ such that }\underset{w \in I}{ \bigcup} \block_w  \cap \view \subseteq \secret \text{ and  there is no such }I' \text{ with } I'\supset I. 
\end{equation*}
\end{definition}
\begin{example}\label{Exa:SecConfig}
In line~8 in the program of Example~\ref{Exa:Prog} the value of $x_5$ is declassified to $x_6$, being prepared by the FlowTracker's computation of the 
temporary view for $x_5$, the partition in the grayly shaded area of Figure~\ref{Fig:Partition}.
For this declassification, the CIECensor determines the security configuration secret-wise for the policy in Example~\ref{Exa:Scenario}.
\begin{itemize}
\item $\secret_1 = \{(id,\mathit{A},\mathit{B},c_1) \mid \mathit{A}\in \mathit{dom}(A), \mathit{B}\in\mathit{dom}(B) \}$:\\
There is no block in the partition for $x_5$ completely contained in $\secret_1$.
\item $\secret_2 = \{(id,\mathit{A},\mathit{B},c_2) \mid \mathit{A}\in \mathit{dom}(A), \mathit{B}\in\mathit{dom}(B) \}$:\\
The blocks indexed by $(\mathit{A},c_2)$ with $\mathit{A} \in \mathit{dom}(A)$, in the right upper corner of the diagram, each are 
contained in $\secret_2$ and even so is their union.
\item $\secret_3 = \{(id,\mathit{A},b_2,c_3) \mid \mathit{A}\in \mathit{dom}(A) \}$:\\
There is no block contained in $\secret_3$, but each of the blocks\\$\{(id,\mathit{A},b_2,c_1),(id,\mathit{A},b_2,c_3)\}$ with $\mathit{A}\in\mathit{dom}(A)$ consists only of states containing confidential information as defined by $\secret_1$ and $\secret_3$ together.
\end{itemize}
Gathering these results, we obtain the security configuration $\secconf = (\genRange' , \{ \{(\mathit{A},c_2) \mid \mathit{A} \in \mathit{dom}(A)\} \})$ where its domain $\genRange'$ is the set of all possible block indices, i.e., $\{(\mathit{A},\mathit{B}) \mid \mathit{A} \in \mathit{dom}(A), \mathit{B}\in\mathit{dom}(B)\}\cup \{(\mathit{A},c_2),(\mathit{A},c_4) \mid \mathit{A}\in\mathit{dom}(A)\}$.

Somehow counter-intuitively, the block indices $(\mathit{A},b_2)$ with $\mathit{A}\in\mathit{dom}(A)$ are not to be found among the violating sets of $\secconf$ because the knowledge represented by each of the indexed blocks includes that one of $\secret_1$ and $\secret_3$ holds, but not which of them.
If the CIECensor included these indices, it might mistakenly detect a policy violation. 
Due to the security configuration being non empty we conclude that there is a potential policy violation so that the censor must possibly generalize the value of $x_5$ before its declassification.
\end{example}

The meaning of a set $I$ in a security configuration, considering the context of a declassification $\icontrol{x_{src}}{x_{dest}}$, is 
that knowing the value of $x_{src}$ is in $I$ enables the partner to infer that a confidential piece of information $\secret$ is contained in the actual abstract information state.  
In this sense, a set $I'' \subset I$ stands for more specific knowledge which is harmful if $I$ is, so that it is omitted from the respective security configuration.

\begin{example}
Continuing the previous example, we see that from knowing that the value of $x_5$ is in $I=\{(\mathit{A},c_2) \mid \mathit{A} \in \mathit{dom}(A)\}$ the partner might conclude that the actual abstract information state is in the union of the respective blocks, viz. $\underset{(\mathit{A},c_2)\in I}{\bigcup} \{(id,\mathit{A},\mathit{B},c_2) \mid \mathit{B} \in \mathit{dom}(B)\}$, and hence that the secret $\secret_2$ holds.
Each of the blocks  in this union represents more specific knowledge which as such leads the partner to the same conclusion.
\end{example}	

With simple set-theoretic arguments we obtain a more concise declaration of the security configuration.
\begin{proposition}\label{Prop:SecConfig}
The security configuration of Definition~\ref{Def:SecConfig} can be rewritten to\\
$(\genRange',\{ I_\secret  \mid \secret \in \conf \text{ with } I_\secret=\{w \in R' \mid  \block_w  \cap \view \subseteq \secret \}\text{ and }I_\secret\neq\emptyset\})$.
\end{proposition}
To determine the configuration algorithmically, for each secret in $\conf$ we need to iterate through $R'$ instead of searching $\pset{R'}$ for the
maximal sets $I$ described by Definition~\ref{Def:SecConfig}.\\[1ex]
	
\noindent\textbf{8. Filtering and modifying by generalization}.
	% to avoid meta-inferences}.
	If the CIECensor on checking the security configuration finds it not empty, which indicates a possible policy violation,  the  CIECensor 
	%distinguishes
	considers whether there would be  an \textit{actual violation}. 
	% or not.
	Clearly, this is the case if the block 
	%formed by intersecting the actual block of the tentative addition 
	that corresponds to the actual value, 
	combined with the previous view, 
	%and the actual block of the previous view 
	is contained in a policy element,
	and thus
	% the corresponding actual value
	this value may not be revealed. However, to avoid meta-inferences such a hiding has to be made \textit{indistinguishable} from
	the treatment of at  least one different value. Accordingly, the \textit{CIECensor} has to apply a precomputed \textit{distortion table},
	as exemplified in~\cite{BiWe08}, 
	that (i) clusters possible values such that the union of their blocks is not contained in any policy element 
	and (ii) determines for each cluster a suitably \textit{generalized value}, similarly as for \textit{k}-anonymity.
	
\begin{definition}[Distortion table]\label{Def:DistTab}
Let $\genRange$ be a finite subset of $\vals$.
Then a  distortion table over $\genRange$ is a function $\disttab: ( \{\genRange' \in \pset{\genRange}\} \times \pset{\pset{\genRange'} \setminus \genRange'}) \times \genRange \rightarrow \genRange$ such that 
\begin{enumerate}
\item for all $\genRange' \in \pset{\genRange},v\in \genRange$ it holds $\disttab ((\genRange',\emptyset) , v ) = v$;
\item for all row indices $ri$ of the form $(\genRange', \vioSets ) \in  \pset{\genRange} \times \pset{\pset{\genRange'}\setminus \{\genRange'\}}$ and $v \in \genRange$\\such that $\{w \in \genRange |    \disttab ( ri , w ) = v \} \cap \genRange' \neq \emptyset$\\there does not exist $I \in \vioSets$ such that 
 $\{w \in \genRange |    \disttab ( ri , w ) = v \} \cap \genRange' \subseteq  I.$
\end{enumerate} 
\end{definition}
	
	Applying the distortion table then means that the \textit{CIECensor}, after it has determined a security configuration $\secconf$ of Definition~\ref{Def:SecConfig}, looks up the generalized value of $v \in \genRange$ as $g=\disttab (\secconf , v)$.
All values $w$ with  $\disttab(\secconf , w)=g$ form a cluster as stated by (i) above Definition~\ref{Def:DistTab} and ensured by the definition of the distortion table and the security configuration.
	After generalization, the CIECensor 	updates the previous view with the partition derived from the clustering
	and returns the generalized value to the destination variable of the declassification assignment.

\begin{example}\label{Exa:DistTab}
Reconsider the security configuration 
$(\genRange',\{ \{(\mathit{A},c_2) \mid \mathit{A} \in \mathit{dom}(A)\} \})$ from Example~\ref{Exa:SecConfig} denoted by $\secconf_1$.
The CIECensor determines this configuration for the declassification of $x_5$.
For this configuration a distortion table may contain the following row\\
\begin{center}
\begin{small}
\begin{tabular}{ | p{10ex} @{\hspace*{1ex}} || p{8ex} *{7}{| p{8ex}} |}
\hline
 security configuration  &  \multicolumn{8}{l |}{value to be declassified}\\
 & \ldots & $(a_n,b_m)$ &  $(a_1,c_2)$ &  \ldots &  $(a_n,c_2)$ & $(a_1,c_4)$ & \ldots &$(a_n,c_4)$ \\\hline\hline 
 $\secconf_1$ & \ldots &  $(a_n,b_m)$ & $(a_1,g_C)$ & \ldots & $(a_n,g_C)$ & $(a_1,g_C)$ & \ldots & $(a_n,g_C)$ \\ 
\end{tabular}
\end{small}
\end{center}

\noindent If the value of $x_5$ was $(a_1,c_4)$, which is actually harmless, the CIECensor would generalize it to $(a_1,g_C)$ to hide the harmful value of $(a_1,c_2)$ behind the same output. 
Afterward, it would set the previous view to $\{ (id,a_1,\mathit{B},c_2),$ $(id,a_1,\mathit{B},c_4) \mid \mathit{B}\in\mathit{dom}(B)\}$.
The CIECensor returns other harmless values ungeneralized such as $(a_n,b_m)$.
\end{example}
Section~\ref{Sec:DistTab} will present an example for a method to compute the generalization of a value given the security configuration without explicitly storing such a table, but a generalization hierarchy over $\genRange$ as inspired by generalization schemes for \textit{k}-anonymity~\cite{FuWaChYu10}.

	As part of the \textit{programming discipline},  the set of possible return values of a basic information reaction and the set of (crucial) values of external reactions 
	(depending on the abstract information state) has to be kept suitably small to manage the computation of the distortion table.
	Here, we greatly simplify this challenge by requiring that such a finite set $\genRange$ of crucial values indeed exists.
  To enable further processing of generalized values within the \textit{open realm},
	all operators used in the program have to be suitably \textit{redefined} by overloading. 
	However, we will abstract from this aspect in this work.

	\section{Implementing and specifying the mediator framework}\label{Sec:Specifi}
	In this section, by a Java-based exemplary instantiation of the framework we will first illustrate how all parts of the framework are to be combined
	and work together to realize the desired mediator.
	In particular, we illustrate how the mediator synchronizes program execution with dynamic tracking by the FlowTracker and declassification by the CIECensor. 
	Moreover, to prepare for a formal verification of the framework, we will describe the mediator's behavior by runs of the considered system $\system$ as introduced in the scenario description of Section~\ref{Sec:Overview}. 
Based on the mediator-side run specification, we will then consider the cooperation partner's capabilities as a worst case as those of a more powerful observer whose behavior we will define as part of these runs.
In particular,  we consider the partner's reasoning as the observer's knowledge operator $\K$ which we will define in Section~\ref{Sec:Observer}. 
	
\subsection{Implementation in Java using Paragon}\label{Sub:Impl}
\begin{figure}
	 \centering
  \resizebox{0.95\textwidth}{!} {%
    \def\svgwidth{\textwidth}
    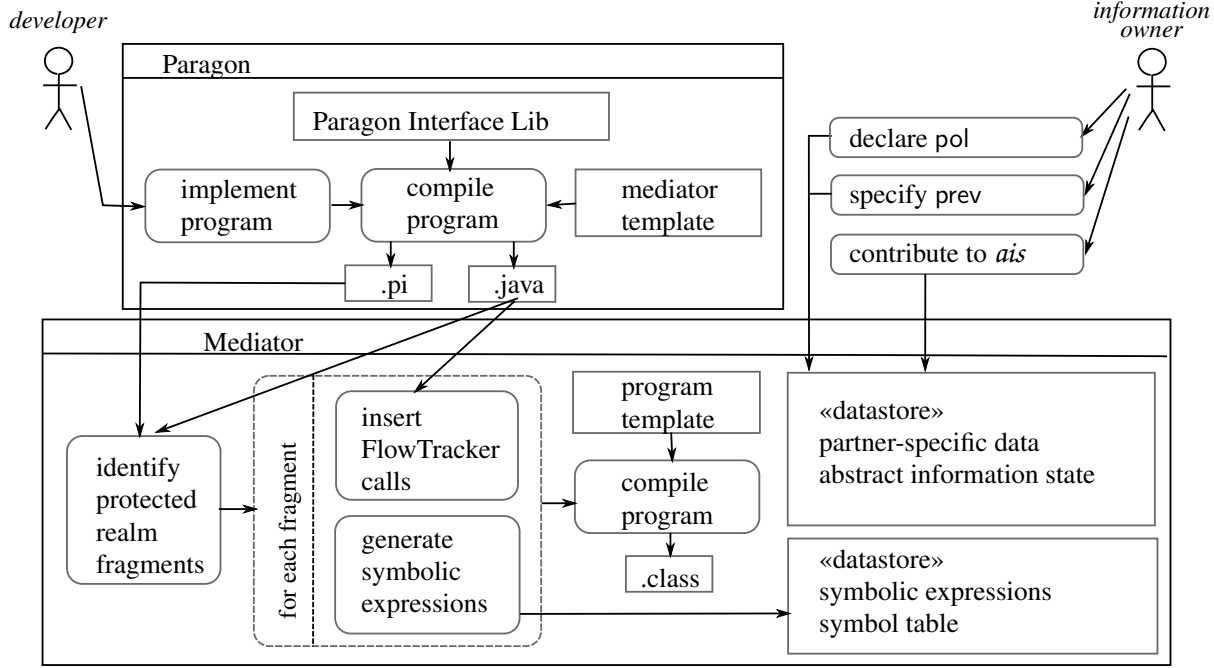
  }
\caption{Instantiation of the mediator framework in a Java programming environment}\label{Fig:Instantiation}
	\end{figure}

\begin{figure}
	\vspace*{3ex}
	\def\svgwidth{\textwidth}
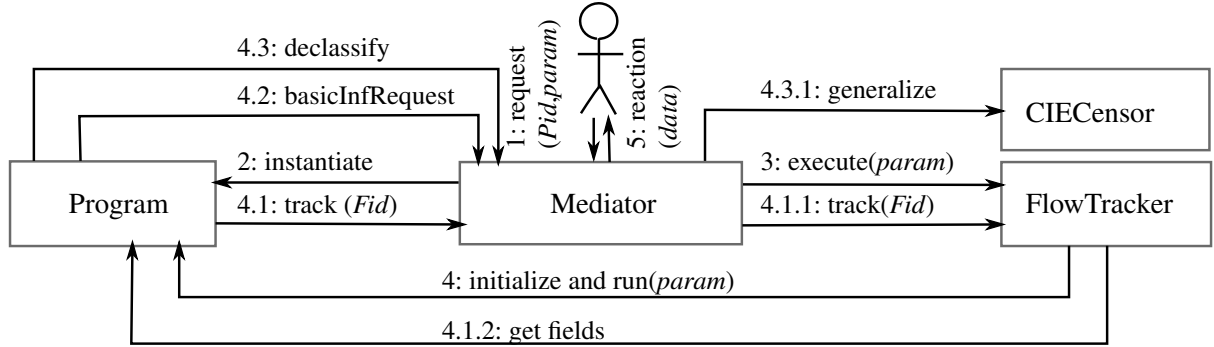
\caption{Communication among components of the instantiated framework}\label{Fig:Communication}
	\end{figure}

A leading idea behind the design of the framework is to base on
established technologies such as language-based information flow control and symbolic execution
in order to facilitate a holistic control as anticipated in Section~\ref{Sec:Overview}.
In this subsection, we overview a Java-based implementation of the framework, conducted in a master thesis~\cite{Za16}.
This overview first  demonstrates along Figure~\ref{Fig:Instantiation} how
to instantiate the framework with a basic utilization of existing program libraries including such for the mentioned technologies.
This instantiation reveals challenges incurred in any implementation of the framework and suggests respective, preliminary solutions.
Second, this overview demonstrates along Figure~\ref{Fig:Communication}  how the mediator's components generally work together.
 The aim of their joint functioning is to ensure confidentiality as stated in Property~\ref{Def:Conf}.
 As a basis for its verification, we finally formalize the framework as a system of runs in the next two subsections.

\noindent\textbf{Instantiation}.
As a first major task, a \emph{developer} implements a program for the intended service of the mediator in a programming environment
which consists of the \emph{mediator template} with the abstract-information-program interface and libraries of .pi files for the Java-based language \emph{Paragon} with its own compiler.  
Supporting the developer's task, this compiler ensures Item~1 (Isolation) and Item~2  (Sharing) by rejecting programs that release information from the abstract information state otherwise than through a designated declassification method. 
To this end,  methods offered by the mediator template are annotated with \emph{flow policies} of Paragon to specify the permitted information flow accordingly
while the PI library  (\textbf{P}aragon \textbf{I}nterface) does likewise for a basic selection of the Java API. 

A challenge regarding the setup of the programming environment is that the Paragon compiler does not support all Java features, see~\cite{BrDeSa13,ParagonWebsite}, and the PI library is limited so that the full framework could only be written in Paragon with much effort if at all. 
Instead the framework is preliminarily reduced to its main functionality, the interface-wise access to the abstract-information state and the interplay of the control components, 
as method stubs in the mediator template. 
A further challenge is to utilize the .pi file from the compilation of the program to  identify fragments processed in the protected realm in an automated manner and to put the execution of such fragments under control of the dynamic monitor, consisting of FlowTracker and CIECensor. 

As a second major task, \emph{the information owner}, or a security engineer on his behalf,  stipulates partner-specific definitions, the confidentiality policy $\conf$ and the partner's background $\view$. 
These are persisted in a \emph{data-store} to enable Item~3 (History-based policy compliance). 

 The further major tasks are performed by compilers designed for the framework in an automated manner, also using Java CC which generates compiler code from a grammar in BNF.   
First, for Item~4 (Flow tracking) fragments of protected realm processing are identified, following the flow policies which are derived by the Paragon compiler from flow policies initially annotated by the developer
and which are then added by the compiler into the .pi file for the program.
To synchronize the FlowTracker with the program, FlowTracker calls are inserted into the code just before each fragment.  
To allow the FlowTracker to read variables of the program, the code of the program is filled into a program template and transformed by substituting program variables with object fields. 
This way we basically deal with the Java-inherent challenge that only the program is allowed access to its method stack where program variable values reside, whereas object fields reside in the object heap which may be principally shared by all threads. 

Second, for flow tracking according to Item~5 (Implicit flows) the manipulation of each high variable by a protected realm fragment
is encoded into a respective symbolic expression.
Symbolic expressions are kept in the data-store with an index on an identifier of the respective fragment together with a symbol table  to look up the initialization of each symbol during the mediator's working phase.
An advantage of shifting the determination of symbolic expressions from the mediator's working phase to compile time is that it may save computational 
effort during the mediator's working phase.
Especially, this should be the case if the compiler tested whether a symbolic path condition is not satisfiable to eliminate the term connected to this path condition. 
Such satisfiable tests involve high computational costs at compile time, but reduce the computational costs for interpreting symbolic expressions
during the working phase. 
In this article, however, such optimizations by satisfiability testing are not further treated.

This task completes the instantiation of the framework which is run with a library for its full functionality in place of the mediator template. 
To realize this functionality, a noteworthy challenge is that for declassifying the value of  a variable, not only the value, but also the variable identifier is needed by the CIECensor to obtain the temporary view of that variable according to Item~6 (Tentative addition) and Item~7 (Harmlessness).
A significant challenge is that generalized primitive data types in Java must be defined and the associated operators overloaded according to Item~9 (Generalization). For our basic instantiation of the framework, we avoided the last challenge by converting such types into Strings whenever data are transferred from the program to framework components. 

\noindent\textbf{Communication among components}.
In our Java-based implementation, the cooperation partner poses his requests, to be handled by the mediator, to an instance of the class Mediator in step 1 of Figure~\ref{Fig:Communication} and the Mediator instance in response initiates the run of a program specified in the request by an identifier \textit{Pid}, e.g. a URL.
Such an identifier may select one among several programs each designed for the processing of a specific request type.
As shown previously in Figure~\ref{Fig:Concept}, the components of the dynamic monitor  are to be isolated from the program to prevent their manipulation by its developer.  
Therefore, the Mediator works as a proxy of these components according to the following standard steps.

 In step~2, the Mediator loads the class Program, as selected by \textit{Pid}, and instantiates it binding to the new instance the symbolic expressions and related symbol tables in the data-store. Then, in step~3, the Mediator hands over execution to the FlowTracker to let it synchronize with program execution. 
Accordingly, in step~4 the FlowTracker sets fields of the Program instance to parameter values from the request and runs the Program. 
During this run, the Program synchronizes with the FlowTracker via its proxy, the Mediator, in step~4.1, right before a protected realm fragment with identifier \textit{Fid} as determined during compilation.
For synchronization, the track message activates the FlowTracker in step~4.1.1 which may access program variables through getters of object fields in step~4.1.2. 
Moreover, during its run, the Program first accesses the abstract information state in step~4.2 and then declassifies values via the Mediator in step~4.3 which hence calls the CIECensor for value generalization in step~4.3.1.

	\subsection{Specification of the mediator}\label{Subsec:Mediator}
	Now, we leave the specific instantiation illustrated in Figure~\ref{Fig:Communication} and provide a general pattern for
	the mediator, including its dynamic monitor, with the aim of achieving and verifying Property~\ref{Def:Conf} (Confidentiality Preservation).
	We give this pattern in form of a run specification, which on the one hand defines the general functioning of the desired mediator.
	In view of the exemplary instantiation, this functioning is the joint functioning of the four classes Mediator, Program, CIECensor and FlowTracker (and thus, in particular, is more comprehensive than the Mediator class which mainly serves as a proxy for isolating the other classes from one another).

	On the other hand, this run specification defines an observer of the mediator's activities with more powerful means than the cooperation partner presumably has
	as an attacker against the confidentiality policy, as  illustrated in Figure~\ref{Fig:Observer} and explained in this section.  
	In Section~\ref{Sec:Declarative}, we will formally prove that the mediator, if instantiated according to pattern of this run specification, achieves Property~\ref{Def:Conf} (Confidentiality Preservation) which has been introduced in Section~\ref{Sec:Overview} and is now formalized by this specification.

The mediator's behavior is defined by runs as functions from time $\mathbb{N}_0$ to its state which consists of 
three  sub-states, namely $\intprocst$ for interaction processing,  $\censorst$  for the CIECensor and $\flowst$ for the Flow Tracker.
In the interaction processing state  $\intprocst = \langle \code  \csep \mem \csep \bstate  \rangle$ the mediator prepares the reaction according to a request from the partner,  executing program $\code$ on the current memory  $\mem$   and  the fixed abstract information state $ \bstate \in \bstates $.
The memory is a  function $\mem: \vars \rightarrow \vals$  and its low projection $\mem_\low$  is the restriction of this function to low variables.

Aiming at a unified approach to control, we do not select a specific programming language, but consider a language  $\intproclang$ as some  (inductively defined) set of well-formed programs, including the empty program $\emptyprog$.
We do not further specify this language, but only assume that it is interpreted by  a one-step semantics defined as a function $\evalprog: \intproclang \times (\vars \rightarrow \vals) \times  \bstates  \rightarrow \intproclang \times (\vars \rightarrow \vals) \times  \bstates$, see~\cite{Nielson1999Principles} for examples.
Moreover, we  assume that every sequence of successive applications of this function converges to a result $( \emptyprog , \mem_n , \bstate )$   in a finite number $n$ of steps (termination assumption), and further that $(\code, \mem , \bstate)$ uniquely determines this result  (determinism assumption).
Lastly, we assume that basic information requests are included in $\intproclang$ as a  command of the simplified form $\basicreq_{para}(x)$, where $para$ is a meaningful parameter,  such as the query kinds ``select'' and ``project'' in Example~\ref{Exa:Prog}, and each of them is interpreted by an evaluation function $\evalprog (\basicreq_{para}) : \vals \times \bstates \rightarrow \genRange$ to the finite subset $\genRange$ of $\vals$.

The CIECensor operates on its state $\censorst =  \langle \view \csep \conf \csep \disttab \rangle$\footnotemark and the FlowTracker on its state $\flowst = \langle   \partitions \csep \ftstatus \csep \ftenv  \rangle$\addtocounter{footnote}{-1}\footnotemark\footnotetext{
with the components: previous view  $\view$, confidentiality policy $\conf$, distortion table $\disttab$, temporary views $\partitions$, status $\ftstatus$, security level inference $\ftenv$.
} as outlined in Section~\ref{Sec:Overview} and summarized in the following in the context of interaction processing.
Table~\ref{Tab:Runs} defines the initialization and then inductively the subsequent state of the mediator in a run $r$ at time $t+1$ according to the first case listed in the table which applies to the precondition given for $r(t)$.
Each such precondition requires a specific form of the components of sub-states of the mediator in $r(t)$.
In the postcondition for $r(t+1)$, only components of the mediator which change are displayed, others are  omitted.

\begin{table}[t!]
\caption{Inductive specification of the mediator in a run $r$ in terms of the precondition and postcondition of the state transition}\label{Tab:Runs}
\begin{center}
\begin{tabular}{|| l | p{1.2cm}  l ||}
\hline
\multicolumn{3}{|| l ||}{Initialization}\\\hline
& \multicolumn{2}{| l ||}{System parameters:}\\
& \multicolumn{2}{| l ||}{$\intprocprog \in \intproclang$ mediator program,}\\  
& \multicolumn{2}{| l ||}{$\bstates$ set of abstract information states,}\\
& \multicolumn{2}{| l ||}{$\conf \subseteq \pset{\bstates}$ confidentiality policy,}\\
& \multicolumn{2}{| l ||}{$\disttab: ( \{\genRange' \in \pset{\genRange}\} \times \pset{\pset{\genRange'} \setminus \genRange'}) \times \genRange \rightarrow \genRange$ distortion table}\\
& \multicolumn{2}{| l ||}{\quad over a fixed, finite range $\genRange$ of values from $\vals$,}\\ 
& \multicolumn{2}{| l ||}{$\env$ security level inference over $\intproclang$,\footnotemark and} \\
& \multicolumn{2}{| l ||}{$\mem^0$ a mapping from $\vars$ to fixed default values and program arguments}\\\hline
$r(0)$& $\intprocst (r,0)$:&  $\code (r,0) = \intprocprog$,   $\mem (r,0) = \mem^0$,  $\bstate (r,0) = \bstate$\\
         &                        &   for some $\bstate \in \bstates$\\
          & $\censorst (r,0)$:&  $ \view (r,0) = \bstates$,  $ \conf (r,0) = \conf$,  $\disttab (r,0) = \disttab$\\
          & $\flowst (r,0)$:&  $\partitions (r,0) =  \partitions^0$ where $\partitions^0 (x) = (\bstates_v)_{v \in \{\mem^0 (x)\}}$\\
          & & for all $x\in\hvars$\\
          & &  $\ftstatus (r,0)  = \ftidle$,  $\ftenv (r,0) =\, \env$\addtocounter{footnote}{-1}\footnotemark\\\hline\hline
\multicolumn{3}{|| l ||}{Tracking of protected realm processing}\\\hline
\multicolumn{3}{|| l ||}{(1) Start flow tracking}\\\hline
$r(t)$  &  $\code(r,t)$ & $\hasform \cm{h}\code ; \cm{l}\code$ where $\cm{h}\code: \high$,  $\cm{l}\code:\low$ and\\ 
&  &  \hspace*{2ex} for any other such sequence $\cm{h}\code';\cm{l}\code'$ the subprogram $\cm{h}\code'$ is a prefix of $\cm{h}\code$ \\
& $\ftstatus(r,t)$ & $\hasform \ftidle$\\
& $\partitions$  & $= \evalsymb (\trans (\cm{h}\code , \env))( \partitions (r,t) , \mem_\low (r,t) )$\addtocounter{footnote}{-1}\footnotemark\\\hline
$r(t+1)$ & $\intprocst$ & $= \evalprog (\seq{ \cm{h}\code;\ftstop ;\cm{l}\code   \csep  \mem (r,t) \csep \bstate (r,t)})$ \\
          & $\flowst$ &  $= \langle \partitions \csep \fttracking  \csep \env \rangle$\addtocounter{footnote}{-1}\footnotemark\\\hline
          \multicolumn{3}{|| l ||}{(2) Stop Flow Tracking}\\\hline
$r(t)$ & $\code (r,t)$ &  $\hasform \ftstop ; \cm{rest}\code $\\\hline
$r(t+1)$  & $\intprocst$ & $ = \seq{\cm{rest}\code \csep \mem (r,t) \csep \bstate (r,t)}$\\
&  $\flowst$ &  $= \langle \partitions (r,t) \csep \ftidle \csep \env \rangle$\addtocounter{footnote}{-1}\footnotemark\\\hline\hline
\multicolumn{3}{|| l ||}{Declassification}\\\hline
\multicolumn{3}{|| l ||}{(3) Generalize value by CIECensor}\\\hline
$r(t)$ & $\code (r,t)$ &  $\hasform \icontrol{x_{src}}{x_{dest}}\cm{; rest}\code$ where $x_{src} \in \hvars$, $x_{dest} \in \lvars$\\
         &  $(\view , g )$ &  $= \censor ( \censorst (r,t) , \partitions (r,t)(x_{src}) , \mem (r,t) ( x_{src}))$\\\hline 
$r(t+1)$ &  $\intprocst$ & $= \langle \cm{rest}\code \csep \mem(r,t)[x_{dest} \mapsto g ] \csep \bstate (r,t) \rangle $ \\
             &  $\censorst$ & $=  \langle \view \csep \conf \csep \disttab \rangle$\\\hline
\multicolumn{3}{|| l ||}{(4) Forward value uncensored}\\\hline
$r(t)$ & $\code (r,t)$ &  $\hasform \icontrol{x_{src}}{x_{dest}}\cm{; rest}\code$\\\hline
$r(t+1)$ & $\intprocst$ &  $= \langle \cm{rest}\code \csep \mem(r,t)[x_{dest} \mapsto \mem(r,t)(x_{src})] \csep \bstate (r,t) \rangle $ \\\hline\hline
\multicolumn{3}{|| l ||}{(5) Interaction processing}\\\hline
$r(t+1)$ & $\intprocst$ &  $= \evalprog (\intprocst (r,t))$\\\hline
\end{tabular}
\end{center}
\end{table}

\footnotetext{
For the sake of simplicity, the notation just indicates the use of an appropriate level inference $\env$, but is not accurate about the representation of the security levels for subprograms and expressions of the mediator program $\intprocprog$.
These levels are usually derived by means of $\env$ during compilation of the complete mediator program.
}

At initialization, the program variables and all fragments of program  $\code$ are assumed to have security levels according to Item~1 (Isolation) and Item~2 (Sharing) of Section~\ref{Sec:Overview} induced by security level  inference $\ftenv (r,t) =\, \env$. 
The not formally stated set $\Gamma$ defines the fundamental security levels of the output for the cooperation partner being low, the parameters to basic information requests being low and the basic information reaction from the abstract information state being high, as we have alluded in Item~1 (Isolation) in Section~\ref{Sec:Overview}.
After initialization, in case (1) (Start flow tracking), if interaction processing reaches a protected realm fragment, here a high-level prefix $\cm{h}\code$ of the rest program $\code (r,t)$, the mediator starts the FlowTracker which changes its status  $\ftstatus$ from \textit{idle} to \textit{tracking}. 
While tracking, the FlowTracker cannot be called again according to the precondition in $r(t)$ for such a call.
This way, the specified runs synchronize program execution and flow tracking  at the start of protected realm fragments like the Java implementation in Section~\ref{Sub:Impl} does.

By means of a still unspecified function $\trans$, the FlowTracker extracts symbolic expressions from the program prefix $\cm{h}\code$, one for  each  high-level  variable manipulated by the prefix.  
This extraction could also take place at compile time as in Section~\ref{Sub:Impl} so that the FlowTracker would only look up the precompiled expressions. 
For each such high variable, while in status \textit{tracking} the FlowTracker interprets these expressions to modify the temporary view $\partitions$ of the variable using a still unspecified function $\evalsymb$.  For this interpretation, the FlowTracker might need the value of low-level program expressions and thus needs access to the program's low memory $\mem_\low$.  
The two still unspecified functions $\trans$ and $\evalsymb$ are introduced in Section~\ref{Sec:FlowTracker}.
After the high-level fragment $\cm{h}\code$, the mediator inserts the command $\ftstop$ which stops the FlowTracker.

In case (2) (Stop flow tracking), the FlowTracker's status is reset to \textit{idle} by the special command $\ftstop$ so that it may be called again for tracking in case (1).
In case (3) (Generalize value by CIECensor), the CIECensor is called as a function $\censor$ for the declassification of a high-level variable and computes a possibly generalized value $g \in \genRange$ and an element in $\pset{\bstates}$ for the previous view. 
The function $\censor$ is introduced in Section~\ref{Sec:Declarative}.
The fourth and fifth case include a special case of declassification and general interaction processing by program execution, respectively. 

\subsection{The observer representing the partner as an attacker}\label{Sec:Observer}
\begin{figure}[t]
	\vspace*{-4,75cm}
	\begin{center}
	\def\svgwidth{\textwidth}
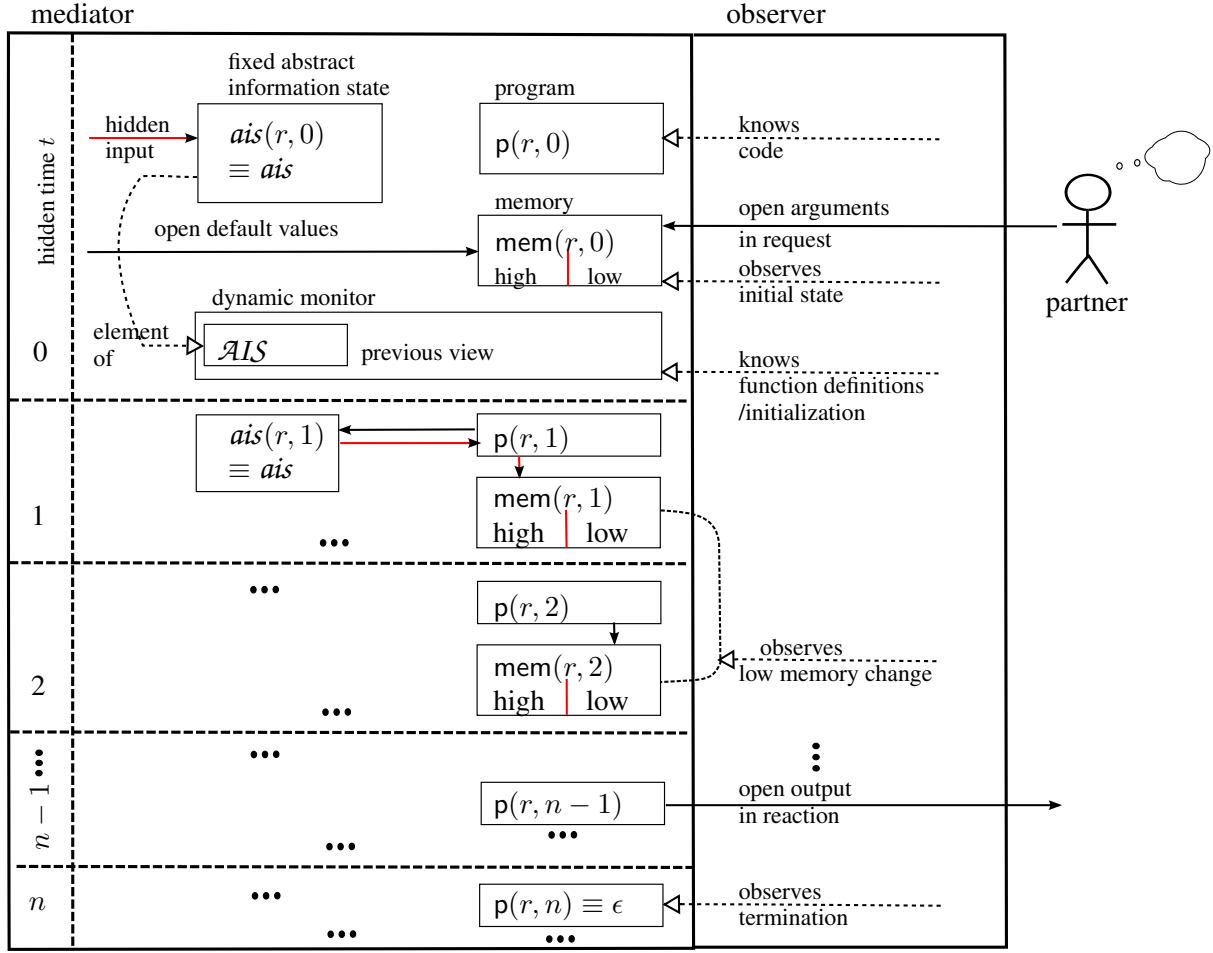
\vspace*{4.75cm}
	\end{center}
	\caption{The mediator's functioning as perceived by the observer in system $\system$, as a worst case scenario for the cooperation partner's options of inference as an attacker against the confidentiality policy}\label{Fig:Observer}
	\end{figure}

As part of the system $\system$ we specify the state transitions of an observer which has  presumably more powerful means than the partner as illustrated  in Figure~\ref{Fig:Observer}.
The observer has  fundamental, initial knowledge to reason about the mediated abstract information state, namely the system parameters as listed in Table~\ref{Tab:Runs} and the system's specification $\system$. 
Note that contrary to other work the observer knows the initial memory.
The initial memory is set to default values and program arguments, but  is not dependent on the mediated abstract information state which is the only target of the confidentiality policy in this article.
Moreover,  the observer might perceive the mediator's behavior according to the transition function of its state $\obsst$ in a run $r$ as defined in Table~\ref{Tab:RunsObs}.

Based on the mediator-side specification of a run $r$, this table defines a sequence of observable events $\obsst(r,t+1)$ inductively for the mediator's internal time $t$ as sketched in Figure~\ref{Fig:Observer}.
Starting with the observation of initial memory,  inductively the next possible observation at time $t+1$ is appended to $\obsst(r,t)$ according to the first case listed of which the precondition on the mediator's state holds in $r(t)$. 
More specifically, the observer is informed about progress of computation by the mediator through changes of low memory in case (1)  or termination in case (2), including completion of the reaction stored in a low variable and  sent to the partner.
But neither is the observer informed about changes in high memory nor about the mediator's internal clock, the time $t$.

\begin{table}[t!]
\caption{Inductive specification of the observer in a run $r$ in terms of the precondition on the mediator's state and the postcondition for the next observed event}\label{Tab:RunsObs}
\begin{center}
\begin{tabular}{|| l |  p{1.5cm} l ||}
\hline\hline
\multicolumn{3}{|| l ||}{Initialization}\\\hline
$r(0)$  & $\obsst(r,0)$ & $= \mem (r,0)$  \\\hline\hline
\multicolumn{3}{|| l ||}{(1) Low memory change}\\\hline
$r(t)$   &  \multicolumn{2}{| l ||}{the active command in $\code (r,t)$ assigns $v$ to $x \in \lvars$}\\
           &  \multicolumn{2}{| l ||}{with $v \neq \mem (r,t)(x)$}\\\hline
$r(t+1)$ & $\obsst(r,t+1)$ & $= \obsst (r,t).(x,v)$\\\hline\hline
\multicolumn{3}{|| l ||}{(2) Termination}\\\hline
$r(t)$ &  $\code(r,t)$ & $\hasform \emptyprog$\\\hline
$r(t+1)$ & $\obsst(r,t+1)$ & $= \obsst (r,t).\cm{End}$\\\hline\hline
\multicolumn{3}{|| l ||}{(3) No observation}\\\hline
$r(t+1)$ & $\obsst(r,t+1)$ & $=\obsst (r, t)$\\\hline\hline
\end{tabular}
\end{center}
\end{table}

However, being rational the observer could infer information about high memory changes, or the mediator's internal time, and about the abstract information state. 
We suspect that the observer, being curious and a semi-honest attacker against the policy, intends to infer pieces of confidential information being contained in this state. 
These inferences are derived by the knowledge operator $\K$ on $\bstates$ defined by $\obsst$ in $\system$, analogously as for example in~\cite{BaNaRo08},
\begin{multline}
\label{Def:Knowl} \K (r, t) = \{ \bstate \in \bstates  \mid   \text{exists run } r' \in  \system \text{ and time } t' \in \mathbb{N}_0 \text{ such that }\\
 \bstate (r',0) = \bstate   \text{ and } \obsst(r',t') = \obsst (r,t) \}.
\end{multline}
Applying its initial knowledge about the system, the observer can consider all runs which agree with the observed sequence of events, and rule out the others as possible runs. 
By such considerations, the observer may also gain information about high memory and the mediator's internal time, but they are not the target of
its reasoning as an attacker against the confidentiality policy.

By ruling out possibilities, the observer can narrow down the set of candidates for the actual abstract information state after each observation and does not forget previous observations so that it possibly extends its knowledge over the time, i.e., for $t_{\mathit{after}}>t_{\mathit{before}}$ it holds $\K (r,t_{\mathit{after}}) \subseteq \K (r,t_{\mathit{before}})$ (monotonicity of knowledge).
We may understand  the confidentiality requirement $\K (r,t) \not \subseteq \secret$ for each $\secret \in \conf$ in Property~\ref{Def:Conf} now as follows:  the partner as the observer should be sure about the possibility of the confidential piece $\secret$ of information not being contained in the actual abstract information state.

 \section{Confidentiality enforcement by the CIECensor}\label{Sec:Declarative}
  First of all, by way of example we detail the actions taken by the CIECensor according Algorithm~\ref{Algo:Censor} to follow the observer's inferences and to block them.
  After that we discuss and prove that these inferences are effectively blocked by design of the CIECensor and the distortion table 
  if declassification is the only means to transfer information from a variable $x_{src}$ in the protected realm to a variable $x_{dest}$ in the open realm (gradual release as Property~\ref{Def:Gradual}) and the temporary view  of $x_{src}$ provided by the FlowTracker has the intended semantics.
  
\begin{algorithm}[t!]
\caption{CIECensor}\label{Algo:Censor}
\begin{algorithmic}
\REQUIRE CIECensor state $\langle \view  \csep \conf \csep \disttab \rangle$, partition   $ (\block_w )_{w \in \genRange'}$, value $v\in\genRange$
\ENSURE View  $\cm{view} \subseteq \bstates$, value  $g \in \genRange$
\COMMENT{\hspace*{-10pt}Actions:}
\STATE Determine security configuration $\secconf$ by Definition~\ref{Def:SecConfig} using $(\block_w)_{w \in \genRange'}$, $\view$ and $\conf$
\STATE Set $g := \disttab (\secconf , v )$
\STATE Set $\cm{view} : = \view \cap \infintbel ( (\block_w)_{w \in \genRange'} ,\disttab , \secconf , g)$ to be defined in~\eqref{Equ:InfAbstInf} below
\end{algorithmic}
\end{algorithm}
  
\begin{example}\label{Ex:DistortionTable}
First, the CIECensor, summarized in Algorithm~\ref{Algo:Censor}, determines the pertinent security configuration of Definition~\ref{Def:SecConfig}.
To do so, it principally considers all $I \subset \genRange'$ and tests whether the observer  knowing that $x_{src}$ has a value in $I$ could infer a confidential piece $\secret$ of information. 
Among all sets $I$ with a positive test the CIECensor determines maximal supersets and collects them in the security configuration $\secconf$ of Definition~\ref{Def:SecConfig}.

Then, the CIECensor possibly generalizes the declassified value $v$, using the previously computed security configuration, here  $\secconf_1= (\genRange' , \{ \{ (\mathit{A},c_2) \mid \mathit{A} \in \mathit{dom}(A) \}\} )$  from Example~\ref{Exa:SecConfig}.
The means of generalization is a distortion table of Definition~\ref{Def:DistTab} which given the observer's inference options as represented by $\secconf_1$ 
blocks them by appropriate generalization. 
To illustrate this blocking, we reconsider the distortion table from Example~\ref{Exa:DistTab}, repeated below for convenience, on the row index $\secconf_1$.
With this table, by a lookup in the row indexed by the security configuration the CIECensor generalizes the value $(a_1,c_2)$ to the value $(a_1,g_C)$ as  it would do with $(a_1,c_4)$.

\begin{center}
\begin{small}
\begin{tabular}{ | p{10ex} @{\hspace*{1ex}} || p{8ex} *{7}{| p{8ex}} |}
\hline
 security configuration  &  \multicolumn{8}{l |}{value to be declassified}\\
 & \ldots & $(a_n,b_m)$ &  $(a_1,c_2)$ &  \ldots &  $(a_n,c_2)$ & $(a_1,c_4)$ & \ldots &$(a_n,c_4)$ \\\hline\hline 
 $\secconf_1$ & \ldots &  $(a_n,b_m)$ & $(a_1,g_C)$ & \ldots & $(a_n,g_C)$ & $(a_1,g_C)$ & \ldots & $(a_n,g_C)$ \\ 
\end{tabular}
\end{small}
\end{center}

\noindent If the observer knows the security configuration and observes the value $(a_1,g_C)$ for $x_6$ in line~8 in the code of Example~\ref{Exa:Prog}, then it can infer that the original value of $x_5$ was one of $\{(a_1,c_2), (a_1,c_4)\}$, reconstructing the preimage of $(a_1,g_C)$ in the table in row $\secconf_1$.
This inference is harmless since that inferred set of possible values is not contained in the only violating set of  $\secconf_1$. 
By construction, this violating set and all of its subsets are the sole sets of possible values for $x_5$ whose disclosure reveals a confidential piece of information. 

Finally,  the CIECensor adds to the previous view $\view$ the observer's knowledge gained by disclosure of the generalized value.
This knowledge gain is being investigated in the following.
\end{example}
 
Which knowledge the observer could gain about the abstract information state from a single value observed as a low memory change of a variable $x$ is represented by an $\genRange$-indexed partition $\partition$ for that variable. 
Knowing the value $w\in\genRange'$ of $x$, the observer could know that this state is in the set $\block_w$.
Evidently, the observer could gain knowledge from an inference of possible values for the source variable in a declassification,
but usually less precise knowledge than from a single, definite value. 
As illustrated in Example~\ref{Ex:DistortionTable},  the inferred set of possible values should be the preimage of
the generalized value $g$ and entails the knowledge that takes into account all possibilities and is represented by the following set of \textit{inferred abstract information} states.
\begin{equation}\label{Equ:InfAbstInf}
\infintbel (\partition , \disttab , \secconf , g ) =  \underset{w \in \genRange' :  \disttab(\secconf , w ) = g }{\bigcup}  \block_w
\end{equation}
But that this set indeed represents the observer's knowledge gain from value $g$ needs three further steps of justification.

In the first step, the FlowTracker must be designed and verified to compute temporary views correctly according to  the intended meaning of a block $\block_w$, at least at the time of declassification.
\begin{property}[Correctness of flow tracking]\label{Def:Correctness}
Let $r$ be a run with abstract information state $\bstate$ and $t\in\mathbb{N}_0$ a time such that $\code (r,t) \hasform \icontrol{x_{src}}{x_{dest}};\cm{rest}\code$ where $\cm{rest}\code$ might be empty.
If $x_{src} \in \hvars$, then  it holds 
\begin{enumerate}
\item $x_{src}$ is in the domain of $\partitions (r,t)$,
\item the blocks $\partition = \partitions (r,t) (x_{src})$ form a partition covering $\K (r,t)$,
\item $\mem(r,t)(x_{src})=v\text{ iff }\bstate \in \block_v \text{ in } \partitions (r,t)(x_{src}).$
\end{enumerate}
\end{property}
\begin{example}\label{Exa:CorrectFT}
Let us overview the situation of the FlowTracker in Example~\ref{Exa:TempView} which illustrates temporary view computation for the high variable $x_5$ of the program in Example~\ref{Exa:Prog}. 
Take as the time $t$ the program line's number which is being executed.

The observer's  initial knowledge $\K( r,0)$ is $\bstates$ in Example~\ref{Exa:AIS}, independently of the actual run $r$, and does not change during $r$ until declassification in line~8 because until that line low memory is not affected. Each of these runs corresponds to one execution of the program until line~7 on arguments $c_1$ and $c_3$ based on  a different state in $\bstates$.
In particular, there are $\mid \bstates \mid = \mathit{dom}(A) \cdot \mathit{dom}(B) \cdot \mathit{dom}(C)$ many such runs.

In Example~\ref{Exa:TempView}, the temporary view for $x_5$ in line~8 is the partition shown in the grayly shaded area in Figure~\ref{Fig:Partition}. 
First,  if we go through that example again, we notice that the FlowTracker's evaluation yielding this partition does not depend on the actual abstract information state, but on the values $c_1$ and $c_3$. 
Moreover, we notice that the partition covers $\K (r,7) = \bstates = \{(id,\mathit{A},\mathit{B},\mathit{C}) \mid \mathit{A} \in \mathit{dom}(A), \mathit{B} \in \mathit{dom}(B), \mathit{C} \in \mathit{dom}(C)=\{c_1,c_2,c_3,c_4\}   \}$ as required by Property~\ref{Def:Correctness} in its second point.

For the third point of Property~\ref{Def:Correctness}, let us consider the state $\bstate = (id, a_1 , b_2 , c_1)$.  On this state the if-branch is taken and finally in line~7 the value of $x_5$
becomes $(a_1,b_2)$.  In the partition, shown in Example~\ref{Exa:TempView}, this value points to the block $\{(id,a_1,b_2,c_1),$ $(id,a_1,b_2,c_3)\}$ which indeed contains $\bstate$  (left lower corner of the diagram).  Thus, the block indexed by the actual value contains the actual information state.
On the same way backwards,  the block containing $\bstate$ can be found out. Then, its index equals the actual value of $x_5$. 
These two ways of correctly  determining the block from the value looked up in memory,  and the value from the index of the block looked up for $\bstate$ in the partition, respectively, 
correspond to the two implications of the equivalence in the third point of Property~\ref{Def:Correctness}.
\end{example}

In the second step, we should design an algorithm for temporary view computations by the FlowTracker  in a way that does not enable the observer to extend its knowledge through these computations, and should verify this property.
Such an extension of knowledge would be difficult to determine for checking policy compliance. 
These computations can only be observed through their effects on value generalization during declassification, as they are otherwise isolated from interaction processing as specified in Table~\ref{Tab:Runs} and illustrated in Figure~\ref{Fig:Concept}.
Therefore, we require that in every run $r$ the usage of the temporary view of $x_{src}$ during declassification of $x_{src}$ at time $t$ will still let the observer consider a run $r'$ as an alternative run of $r$,  if the observer does so before the declassification at time $t$ due to $\lequiv{(r,t)}{(r',t')}$.
The simplest way to do so is to require that the FlowTracker computes the same temporary views in $r'$ as it does  in $r$.

\begin{property}[Non-interference of flow tracking]\label{Def:NonInterference}
Let $r,r'$ be runs, $t, t' \in \mathbb{N}_0$ times such that  $\code (r,t) \hasform \code(r',t') \hasform \icontrol{x_{src}}{x_{dest}};\cm{rest}\code$, where $\cm{rest}\code$ might be empty, 
 and  $\lequiv{(r,t)}{(r',t')}$.
Then, it holds
\begin{itemize}
\item  the domains  of  $\partitions(r,t)$ and $\partitions (r',t')$ are both $\hvars$
\item and $\partitions (r,t) (x)= \partitions (r',t') (x)$ for all $x \in \hvars$.
\end{itemize}
\end{property}
\begin{example}\label{Exa:NonInterfFT}
We are considering the observer's knowledge gain from the usage of the temporary view of $x_5$ in line~8 of the program of Example~\ref{Exa:Prog}
during declassification.
The observer at the start of a run may observe memory initialization in particular with the  program arguments $c_1$ and $c_3$. 
The system $\system$, specified by Table~\ref{Tab:Runs} and Table~\ref{Tab:RunsObs}, is set up with these parameters among others such as $\bstates$ so that it holds $\obsst (r,0) = \obsst (r',0)$ for all $r, r' \in \system$.
As until program line~8 with the declassification neither a low memory change nor a termination could be observed, it even holds $\obsst (r,t) = \obsst (r',t')$ for all $r, r' \in \system$ and for all $0 \le t,t' \le 8$.
In every run $r$ for the considered program, according to Table~\ref{Tab:Runs}, the FlowTracker works in three subsequent phases of tracking, which each contributes to the computation of $\partitions (r,8) (x_5)$, the temporary view of $x_5$ used for declassification of $x_5$ in line~8: the first phase during the assignment in line~1 (as marked by the underlining), the second phase during the assignment in line~3 and the third and final phase during the sequence from line~5 until line~7. 
In each run, the final result of the temporary view for $x_5$ is the same as in every other run at time $7$, and importantly at time $8$ of declassification
as Example~\ref{Exa:TempView} has illustrated.  
Therefore, Property~\ref{Def:NonInterference} is satisfied in the discussed scenario.
\end{example}
In Section~\ref{Sec:FlowTracker},  according to these two steps and the respective two declarative properties the temporary view computations of the FlowTracker will be detailed, based on an algorithmic approach similar to symbolic execution.

Finally, in the third step, taking  Property~\ref{Def:Correctness} and Property~\ref{Def:NonInterference} together,  we are ready to prove that from a generalized value $g \in \genRange$ the observer gains the knowledge described by~\eqref{Equ:InfAbstInf} and 
 nothing else if the temporary view of the source variable is $\partition$ and harmful inferences of possible values are represented by $\secconf$. 
 This claim is formally stated by the next theorem.
Additionally, for its proof, we need correspondence relations between runs, defined in~\cite{BaNaRo08} and outlined in the appendix, by means of which declassification assignments in a pair of runs can be matched.
Such correspondence relations may be constructed if the security level inference system adheres to the two rules,  no-write-down and no-read-up  (without declassification), 
sketched in Item~1 (Isolation) of Section~\ref{Sec:Overview}.
\begin{theorem}[Declassification under generalization]\label{Thm:Declassification}
Let $\system$ be as specified in Table~\ref{Tab:Runs} and Table~\ref{Tab:RunsObs}, based on a security level inference system guaranteeing Property~\ref{Def:Gradual} (Gradual release) and Property~\ref{Def:NoReadUp} (No-read-up) and Property~\ref{Def:NoWriteDown} (No-write-down), given in the appendix. Moreover, let the CIECensor of $\system$ use Algorithm~\ref{Algo:Censor} and the FlowTracker of $\system$ satisfy Property~\ref{Def:Correctness} (Correctness) and Property~\ref{Def:NonInterference} (Non-interference).  
 Let $r\in\system$ be a run and $t$ a time such that $\code (r,t) \hasform \icontrol{x_{src}}{x_{dest}};\cm{rest}\code$, where $\cm{rest}\code$ might be empty, and $x_{src}\in\hvars$ and $x_{dest} \in \lvars$. 
Moreover, let $ g = \disttab ( \secconf , u )$ with   $ u = \mem (r,t)(x_{src}) \in \genRange $.
Then, it holds $\K (r,t+1) = \K (r, t) \cap \infintbel(\partitions(r,t)(x_{src}), \disttab , \secconf , g).$
\end{theorem}

As a conclusion and summary, we argue and prove formally in Theorem~\ref{Thm:Conf} that by the three actions according to Algorithm~\ref{Algo:Censor} 
the CIECensor explores all the observer's options of inference and blocks them successfully. 
In the first action, the CIECensor by determining the security configuration captures all such options that are given through the disclosure of the generalized value $g$ by declassification.
The CIECensor  identifies these options by inference tests of the form $\block_w \subseteq \secret$ where $\block_w$ is the knowledge gained by a value $w$, which may be generalized to $g$, and $\secret \in \conf$ is a confidential piece of information. 
The form of these tests is suggested by Proposition~\ref{Prop:SecConfig}.
In the second action, the CIECensor employs a distortion table which by design effectively blocks all options of the former kind by appropriate value generalization. 
In the third action,  the CIECensor updates the observer's knowledge in the previous view correctly, as ensured by Theorem~\ref{Thm:Declassification},
and thus captures all options of inference based on the history of observed computation steps performed by the mediator.
Therefore, we can prove confidentiality preservation based on the assumption that the confidentiality policy does not aim to protect what the observer, and hence the presumably less powerful cooperation partner, already initially knows, namely that the abstract information state is in $\bstates$.      
\begin{assumption}\label{Asmp}
For all $\secret \in \conf$ it holds $\secret \neq \bstates$.
\end{assumption}
\begin{theorem}[Confidentiality preservation]\label{Thm:Conf}
 Let $\system$ be as specified in Table~\ref{Tab:Runs} and Table~\ref{Tab:RunsObs} satisfying Property~\ref{Def:Gradual} (Gradual release), and Property~\ref{Def:NoReadUp} (No-read-up) and Property~\ref{Def:NoWriteDown} (No-write-down), given in the appendix.
Let the CIECensor  of $\system$ use Algorithm~\ref{Algo:Censor} and the FlowTracker of $\system$ satisfy Property~\ref{Def:Correctness} (Correctness) and Property~\ref{Def:NonInterference} (Non-interference).
 Suppose further that $\system$ adheres to Assumption~\ref{Asmp}.
Then, $\system$ satisfies Property~\ref{Def:Conf} (Confidentiality preservation).
\end{theorem}

\section{Distortion tables by subtree generalization}\label{Sec:DistTab}
\begin{figure}
\subfloat[Two subtrees $\gtree_1$ and $\gtree_2$ hiding $I$ within the domain $\genRange'$ of all leaves]{\includegraphics[width=6cm]{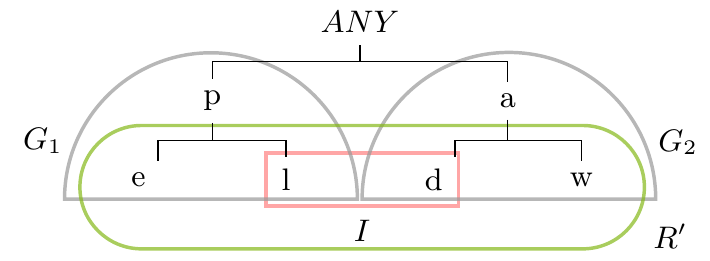}}\hspace*{2ex}
\subfloat[One subtree $\gtree_1$ hiding $I$ within the domain $\genRange'$ of a selection of leaves]{\includegraphics[width=6cm]{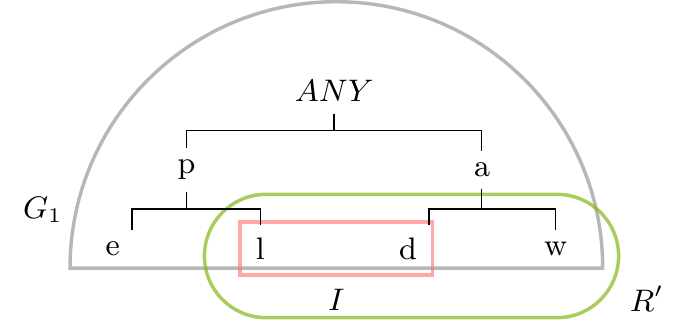}}
\caption{$\secconf$-subtree generalization schemes for a generalization hierarchy over\\$\genRange = \{ \text{p(\textit{rofessional}),e(\textit{ngineer}),l(\textit{awyer}),a(\textit{rtist}),d(\textit{ancer}),w(\textit{riter})}\}$}
\label{Fig:SubtreeGen}
\end{figure}

As it is done for \textit{k}-anonymity, to algorithmically obtain a distortion table we will use an order relation on values in which a value $v_1$ that is above another value $v_2$ in the order is more general than $v_2$, whereas $v_2$ is more specific than $v_1$.
This way, the order should define a generalization hierarchy with a most general top value,  such as the hierarchy shown in Figure~\ref{Fig:SubtreeGen}.
Formally, a generalization hierarchy is a partial order $\taxtree$ on $\genRange \times \genRange$ such that its Hasse diagram is a tree with a designated root  value $\topg$. 
Here, $\genRange$ is again a finite set of values on which the distortion table for the CIECensor must be defined.
As already mentioned in Item~8 (Filtering and modifying) of Section~\ref{Sec:Overview}, we assume that operators of the programming language $\intproclang$ are overloaded for all non-primitive values in $\genRange$,  such as integer intervals as a generalization of integers which is often used for \textit{k}-anonymization.

For an algorithmic definition of a distortion table $\disttab$, the idea is to design a scheme for determining a parent (or self) $g$ of a value $v$ in the generalization hierarchy given a security configuration and to define $\disttab (\secconf , v) = g$. 
This scheme should ensure that the algorithmically defined table meets all requirements of a distortion table in Definition~\ref{Def:DistTab}, and should select a most specific value $g$ for a most informative reaction to the cooperation partner. 
\begin{example}
We set aside our running example for a moment to present the main ideas of the proposed scheme. 
Consider a security configuration $\secconf = (\genRange', \{I\})$ as one of the two shown in subfigures (a) and (b) of  Figure~\ref{Fig:SubtreeGen}.  
A distortion table
\begin{itemize}
\item[(i)]  should cluster values in $\genRange'$ such that none of the clusters is contained in $I$,\\
\item[(ii)] should cover the whole range $\genRange$ of values with possibly several such clusters, pairwise disjoint, and\\
\item[(iii)] should define a generalized value for each cluster.
\end{itemize}
As discussed in Section~\ref{Sec:Declarative}, each such cluster corresponds to the preimage of the respective generalized value, given the row index $\secconf$ of the table.   

An intuitive idea is to cluster \emph{all} values of a subtree in the generalization hierarchy, and generalize all values in this cluster to the root value of the subtree. 
In the example hierarchy on the left,  the values $e, l, p$ of the subtree $\gtree_1$ may be clustered and each of them generalized to $p$.
In literature on \textit{k}-anonymity, the subtree generalization scheme similarly generalizes all leaves of a subtree to its root, instead of generalizing only a selection of its leaves, see e.g. the survey~\cite{FuWaChYu10}.

The task now is to select subtrees as clusters such that all requirements in (i) and (ii) above are met, and to prefer selecting a subtree $\gtree$ contained in another subtree $\gtree'$ over selecting $\gtree'$ because $\gtree$ has a more specific root value than $\gtree'$ has.
In the example hierarchy on the left, selecting $\gtree_1$ is preferred over selecting the complete tree. 

Recapitulating the requirements in (i) and (ii),  we note that an algorithm, which defines a distortion table,  should only explicitly form clusters which contain values of the violating set $I$ 
because only for such clusters the requirement in (i) must be considered. 
Other clusters of values in $\genRange$ should each contain only one respective value which is mapped to itself by the distortion table, producing no information loss at all. 
Such clusters must not be explicitly constructed by an algorithm for the distortion table.

According to the previous considerations, the proposed scheme comprises the following steps.
\begin{itemize}
\item[(I)] Select disjoint subtrees as clusters, which each contains at least one value of $I$,  such that the intersection of a selected subtree with the domain $\genRange'$ of $\secconf$ is not contained in $I$.
\item[(II)] Select as many such subtrees such that $I$ is covered by the union of their values and not any selected subtree $\gtree$  may be replaced  by one of its subtrees $\gtree' \subset \gtree$ without violating the previous requirements. 
\item[(III)] For each selected subtree, generalize all its values to its root, and map each value not contained in any subtree to itself. 
\end{itemize}
In each of the two situations depicted in Figure~\ref{Fig:SubtreeGen},  we may proceed by tentatively selecting each leaf with a value in $I$ as a cluster. 
If a tentatively selected subtree $\gtree$ violates the requirement in (I), we replace it with the subtree rooted in the parent of the root of $\gtree$.
We proceed in this manner until all selected subtrees fulfill the requirement in (I). 
Finally, we remove all subtrees from the selection which are contained in another selected subtree.   
For the situation on the right, we see that neither the leaf $l$, nor the subtree rooted in $p$ fulfill the requirement in (I), so that the complete tree is selected in the process.  
Note that the domain $\genRange'$ of the security configuration must not be limited to the leaf values of the hierarchy as it is in the figure.
\end{example}

Summarizing the above, by a $\secconf$-subtree generalization scheme  we define a selection of subtrees according to the outlined steps (I) and (II)  and 
use this selection, as in step (III), in Algorithm~\ref{Algo:General} to define a distortion table.
\begin{definition}[$\secconf$-subtree generalization scheme]\label{Def:GenScheme}
Let $\genRange$ be a finite subset of $\vals$,  $\taxtree$ a generalization hierarchy over $\genRange$ and $\secconf = (\genRange' , \vioSets) \in \{\genRange' \in \pset{\genRange}\} \times \pset{\pset{\genRange'}\setminus \genRange'}$.  
An $\secconf$-subtree generalization scheme is a set $\gscheme \subset \pset{\genRange}$ of sets $\gtree$ of values such that
\begin{enumerate}
\item every set $\gtree$ in $\gscheme$ forms a subtree of $T$
\item all sets in $\gscheme$ are pairwise disjoint
\item for all $I \in \vioSets$ there exists a selection $S \subseteq \gscheme$ such that 
\begin{enumerate}
\item  $I \subseteq \underset{\gtree \in S }{\bigcup} \gtree$ 
\item and for all $\gtree \in S$ it holds $\gtree \cap \genRange' \not \subseteq I$
\end{enumerate}
\item $\gscheme$ is minimal among all such sets with respect to the order $\gle$ on $\pset{\pset{\genRange}}$
defined by $ \gscheme \gle \gscheme'$ iff for all $\gtree\in\gscheme$ there is $\gtree' \in \gscheme'$ such that $\gtree \subseteq \gtree'$.
\end{enumerate}
\end{definition}

\begin{algorithm}[t!]
\caption{Subtree generalization}\label{Algo:General}
\begin{algorithmic}[1]
\REQUIRE Generalization hierarchy $\taxtree$ over $\genRange$, row index $\secconf \in \{\genRange'  \in \pset{\genRange}\}\times \pset{\pset{\genRange'}\setminus \genRange'}$, value $w\in\genRange$
\ENSURE Generalized value $g \in \genRange$ 
\COMMENT{\hspace*{-16pt}Actions:}
\STATE \label{Algo:DetermineScheme} Determine an $\secconf$-generalization scheme $\gscheme$  of  Definition~\ref{Def:GenScheme}
\IF{\label{Algo:DetermineSubtree} exists $\gtree \in \gscheme$ such that $w \in \gtree$} 
\STATE Set $g$ to the root value of $\gtree$
\ELSE 
\STATE Set $g:=w$
\ENDIF
\end{algorithmic}
\end{algorithm}

We conclude this section by stating and proving that this way we indeed define a distortion table. The proof is in the appendix.
A main step in the proof is that minimality of the selection implies that the selection is unique given the security configuration $\secconf$ and the generalization hierarchy $\taxtree$. 
Hence, Algorithm~\ref{Algo:General} does not need to make a choice in line~1, but is deterministic, so that indeed it computes a function.
\begin{proposition}\label{Prop:General}
Algorithm~\ref{Algo:General} computes a distortion table on $(\{\genRange' \in \pset{\genRange}\} \times \pset{\pset{\genRange'}\setminus \genRange'} ) \times \genRange$ in the sense of Definition~\ref{Def:DistTab}.
\end{proposition}
 
 \section{Operationalizing the FlowTracker}\label{Sec:FlowTracker}
 
The main task of the FlowTracker is to determine the knowledge the observer gains about the abstract information state from a value $w$ of high variable $x$ as this value is finally set after the execution of a high-level piece of program code.
For this task, the FlowTracker inspects all execution paths of the considered program code, each path starting with the same, fixed low memory state, but a different, varied abstract information state.
Through this inspection, the FlowTracker may identify all abstract information states on which the execution of the considered program code results in  the specific value $w$ for  variable $x$.
 These states are collected in a block $\block_w$ of the temporary view for $x$,  so that from the value $w$ of $x$ the observer might only gain the knowledge that the actual abstract information state is in $\block_w$. 
 
 To inspect execution paths of a piece of program code, 
 the FlowTracker translates this code to its execution tree, whose paths correspond to a set of execution paths, using a function $\transet: \intproclang \rightarrow \etlang$. While the programming language $\intproclang$ has been introduced in Section~\ref{Subsec:Mediator}, 
by $\etlang$  we denote the set of all well-formed execution trees including the empty tree $\emptytree$ and defined below.
However, it is not necessary that the FlowTracker inspects all execution paths one-by-one.
Instead, the FlowTracker may abstract from execution paths to  sets of execution paths, so that 
the members of each set are execution paths which result in the same gain of knowledge.  
This way, the FlowTracker aims to inspect the change of temporary views.

In our context, an execution tree should use assignments and branch conditions derived from the translated programming language $\intproclang$,
so that assignments and branch conditions are expressed using  $m$-ary operators from $\Theta_m$, basic information requests from $\basicreqs$ and query functions in $\Omega$ from the language $\intproclang$.
Thereby, each query function tests a designated property of data objects and thus transforms any value taken from $\vals$ to a boolean value $\{\true , \false \}$ as the test result.
\begin{definition}[Execution tree~\cite{BaDaGu12}]\label{Def:ExecutionTree}
An execution tree (ET) is a directed labeled tree $T = ( \nodes , E , C, L ,Start)$ such that
\begin{itemize}
\item $\nodes $ is a set of nodes labeled by assignments (from $\intproclang$),
\item $E \subseteq \nodes \times \nodes$ is a set of control flow edges,
\item $C$ is a set of branch conditions which are boolean expressions from $\intproclang$ as defined by
\begin{align*}
Bool & :=  \true \mid \false \mid  Q ( Op ) \mid  \vars \mid  Bool  \,\mathsf{and}\, Bool \mid  Bool \,\mathsf{or}\, Bool  \mid \\
& \quad\;\; \mathsf{not}\, Bool \mid (Bool) \\
Op &  :=   O_m (Op , \ldots , Op) \mid \vars\\
 Q  & := \Omega  \quad\quad O_m  := \Theta_m\text{ for }m>1, \quad O_1 := \Theta_1 \cup \basicreqs,
\end{align*}
\item $L : E \rightarrow C$ is a mapping from edges to branch conditions,
\item $Start \in \nodes$ is the root node.
\end{itemize}
\end{definition}
 For a meaningful translation of programs, we assume that the one-step semantics of the program is preserved in the sense that for all $\code , \code_1 \in \intproclang$, $\mem , \mem_1 \in (\vars \rightarrow \vals)$, $\bstate  \in \bstates$ such that 
$\evalprog (\code , \mem , \bstate)=(\code_1 , \mem_1 , \bstate)$ it holds $\evalet(\transet(\code)  , \mem , \bstate) = (\transet (\code_1) , \mem_1 , \bstate)$ where execution trees, too, are provided with an appropriate one-step semantics as a function $\evalet$.  For execution trees  $T$ generated this way the termination and determinism properties are inherited from the translated program.

The FlowTracker, however, does not execute the tree on memory and abstract information state, but encodes the manipulation of each high variable by the tree, following each of its paths,  into a respective symbolic expression, in a way similar to symbolic execution, and interprets such an expression as a change of the temporary view for the respective variable.
Symbolic execution is an active, extensive research field and provides many methods for optimization including parallelization and pruning of branches by satisfiability testing of symbolic expressions, e.g.~\cite{Bucur2011ParallelSymbolic,Phan2014AbstractModel} and to be investigated for this framework in the future.

During symbolic execution as defined by Algorithm~\ref{Algo:SE}, the FlowTracker introduces still uninterpreted symbols for basic information reactions  and  low level expressions involved in the manipulation of a high variable using Algorithm~\ref{Algo:TransExpr}.
Symbolic expressions represent alternative paths in dependence of the introduced symbols and in dependence of high variables, indicating branches of paths by $\symBranch$ and their joins by  $\symJoin$, and have the following form:
\begin{align*}
SymExpr & :=     SymExpr   \symJoin  SymExpr  \mid  SymBool  \symBranch SymOp \mid SymOp  \\
SymOp  & :=  O_m (SymOp, \ldots , SymOp ) \mid \symbols \mid \hvars \\
SymBool & := Q(SymOp) \mid  SymBool  \symAnd SymBool \mid SymBool \symOr SymBool \mid\\
           &  \quad \symNot SymBool \mid (SymBool)  \mid SymBool \symBranch SymBool  \mid \symbols \mid \hvars \\
 Q  &  := \Omega  \quad O_m  :=    \Theta_m  \text{ for }m \ge 1
\end{align*}

\begin{algorithm}[t!]
\caption{Symbolic execution $\transsymb$, adapted from~\cite{BaDaGu12}}\label{Algo:SE}
\begin{algorithmic}[1]
\REQUIRE execution tree $T$,  security level inference $\env$ over $\intproclang$  
\ENSURE  symbolic state  $\symb: \hvars \rightarrow  SymExpr$,\\
\quad symbolic initialization function $\syminit: \symbols \rightarrow Bool \cup Op$
\COMMENT{\hspace*{-9pt}Actions:}
\STATE Depth-first traversal of $T$, assigning to each node $n$\\\quad(1) a path condition $pc(n) \in SymBool$ and\\\quad(2) a symbolic state $\symb(n): \hvars \rightarrow SymExpr$
\STATE Initialize $\symb(Start)(x) = x$ for all $x \in \hvars$ 
\COMMENT{Symbolic execution}
\STATE Set $\symb(n):= \symb(p)$ for parent $p$ of $n$ in $T$
\STATE Set $bool$ as the branch condition of the edge from  parent node $p$ to node $n$ in $T$
\STATE Set $symbool := \transsymb (bool , \symb(p) , \syminit , \env )$, determining  $\transsymb$\\\quad by Algorithm~\ref{Algo:TransExpr}, possibly extending the domain of $\syminit$
\STATE Set $pc (n)  := pc (p) \symBranch symbool$
\STATE Define $expr$ as in the assignment $x:= expr$ of node $n$
\STATE \label{AlgLine:SymbAssign} Set $\symb(n)(x):= \transsymb (expr , \symb(p) , \syminit  , \env )$, determining  $\transsymb$\\\quad by Algorithm~\ref{Algo:TransExpr}, possibly extending the domain of $\syminit$
\COMMENT{After the traversal}
\STATE \label{AlgLin:Leaves} Define $\{l_1,\ldots , l_k \}$  as the set  of all leaf nodes  of $T$
\STATE \label{AlgLin:Result} For all $x \in \hvars$ set $\symb (x):= pc(l_1) \symBranch \symb (l_1)(x) \symJoin \ldots \symJoin pc(l_k ) \symBranch \symb (l_k ) (x)$
\end{algorithmic}
\end{algorithm}

\begin{algorithm}[t!]
\caption{Translation $\transsymb$ of expressions to symbolic expressions}\label{Algo:TransExpr}
\begin{algorithmic}
\REQUIRE  $expr \in Bool \cup Op$, symbolic state $\symb: \hvars \rightarrow SymExpr$,\\\emph{global} symbolic initialization function $\syminit: \symbols \rightarrow Bool \cup Op$,\\
security level inference $\env: \vars \rightarrow \{\low , \high \}$
\ENSURE symbolic expression $symexpr \in SymExpr$
\COMMENT{\hspace*{-9pt}Actions:}
\IF{$expr$ contains only variables $x \in \lvars$, but not a basic information request,  or $expr = \basicreq_{para}(x)$}
\STATE Choose symbol $\alpha$ not used in $\syminit$; Set $\syminit(\alpha ):= expr$; Return $\alpha$
\ELSIF{$expr = x\in\hvars$}
\STATE Return  $\symb (x)$
\ELSIF{$expr = \op (expr_1 , \ldots , expr_m )$}
\STATE Return $\op ( \transsymb (expr_1,\symb ,\syminit , \env ), \ldots ,$\\ \hspace*{3cm}$\transsymb (expr_m, \symb , \syminit , \env ))$
\STATE \textbf{else} and so on according to the inductive definition of $expr$ 
\ENDIF
\end{algorithmic}
\end{algorithm}

\begin{figure}
\begin{center}
\includegraphics [width = \textwidth ]{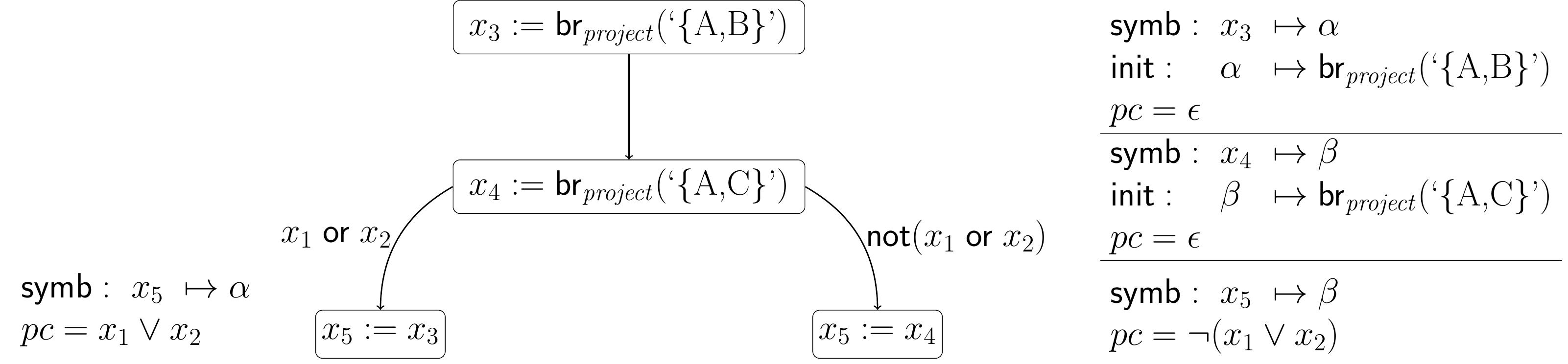}
\end{center}
\caption{Assignments to nodes made by symbolic execution of the high code fragment from line~5 to line~7 in Example~\ref{Exa:Prog} using Algorithm~\ref{Algo:SE} with $\hvars = \{x_1,\ldots , x_5\}$ and $\lvars = \{x_6,x_{rea}\}$}\label{Fig:SE}
\end{figure}

\begin{example}\label{Ex:SE}
We are considering the three phases of flow tracking for the protected realm fragments of the program in Example~\ref{Exa:Prog}, the first being the underlined assignment in line~1,
the second the like in line~3 and the third the underlined sequence from line~5 to line~7.
If activated by the mediator in  case (1) of Table~\ref{Tab:Runs}, the FlowTracker extracts the respective fragment, translates the extracted code to its execution tree and executes Algorithm~\ref{Algo:SE} and Algorithm~\ref{Algo:TransExpr} on that tree. 
Alternatively, the instantiation of the mediator framework overviewed in Section~\ref{Sub:Impl} precomputes symbolic expressions  by a compiler, using Algorithm~\ref{Algo:SE} and Algorithm~\ref{Algo:TransExpr}.

For the first fragment, its execution tree consists only of one node labeled with the assignment in line~1.  
Algorithm~\ref{Algo:SE} processes this node in line~\ref{AlgLine:SymbAssign} by invoking Algorithm~\ref{Algo:TransExpr} which outputs $\symb : x_1 \mapsto \alpha$ and $\syminit: \alpha \mapsto \basicreq_{\mathit{select}}(\text{`C='} + arg_1)$.
In this expression $arg_1$ is a low variable the value of which is concatenated to the string `C='. 
The second fragment is processed analogously.

The execution tree of the third fragment is processed depth-first with Algorithm~\ref{Algo:SE} until line~\ref{AlgLine:SymbAssign} as illustrated by Figure~\ref{Fig:SE}.
In line~\ref{AlgLin:Leaves} and line~\ref{AlgLin:Result}, finally the path conditions and symbolic expressions  of $x_5$ are compiled to the 
result $(x_1 \symOr x_2 ) \symBranch \alpha  \symJoin \symNot (x_1 \symOr x_2) \symBranch \beta $.
\end{example}

To interpret a symbolic expression, the FlowTracker first determines an initial partition for each symbol, called context  $\context: \symbols \rightarrow (\genRange \pfn \pset{\bstates})$,  according to the symbolic initialization function $\syminit: \symbols \rightarrow Bool \cup Op$ as an output of  Algorithm~\ref{Algo:SE}.
To do so, it has to synchronize with the low memory state that interaction processing has produced at the start of flow tracking. 
In the run-based system in Table~\ref{Tab:Runs} we assume that this state is available for the FlowTracker all the time while it is tracking.  
The FlowTracker then uses low memory to determine the argument values for view initialization functions $\initpartition: \basicreqs \times \vals \rightarrow (\genRange \pfn \pset{\bstates})$  of basic information requests taken from $\basicreqs$; and furthermore the FlowTracker uses low memory to interpret the value $v$ determined for a low expression as a single-block partition $(\bstates_w )_{w \in \{v\}}$.

\begin{example}
The symbolic expression $(x_1 \symOr x_2 ) \symBranch \alpha  \symJoin \symNot (x_1 \symOr x_2) \symBranch \beta $ is produced by Algorithm~\ref{Algo:SE} together with the symbolic initialization function $\syminit: \alpha \mapsto \basicreq_\mathit{project}(\text{`\{A,B\}'}),\; \beta \mapsto \basicreq_\mathit{project}(\text{`\{A,C\}'})$ as shown in Example~\ref{Ex:SE}.
The parameters to the basic information requests are string constants, so that the FlowTracker does not need low memory to initialize the partitions for the two symbols $\alpha$ and $\beta$:  the partition for $\alpha$ is shown in Figure~\ref{Fig:Partition} below the dotted crossline  and for $\beta$ above the dotted crossline, respectively.
\end{example}

Then, in the context $\context$,  the FlowTracker interprets composite symbolic expressions $SymExpr$ following their inductive structure as defined by the production rules above. 
In the base case, symbols in $\symbols$ are interpreted by $\context$ and variables $x\in\hvars$ by temporary views $\partitions$.
At the start of flow tracking, temporary views may be available from previous phases of flow tracking. 
Operators in $\Theta$, query functions in $\Omega$ and symbolic boolean operators are all interpreted in an analogous way, exemplified here for $\op \in \Theta_m$ with $\genRange_i \subseteq \genRange$ and $\block_w^i \subseteq \bstates$:
\begin{multline}\label{Eq:SymbIntOp}
\op ( (\block_w^1)_{w \in \genRange_1}, \ldots , (\block_w^m)_{w \in \genRange_m} ) :=  \{ \block_v \mid \block_v  \neq \emptyset \text{ and }
  \block_v = \underset{\op(w_1,\ldots , w_m) = v}{\underset{w_1 \in \genRange_1,\ldots , w_m \in \genRange_m}{\bigcup}} \;\;\underset{i=1,\ldots ,m}{\bigcap} \block_{w_i}^i\}.
\end{multline}
Intuitively, we can understand that the interpretation is correct by viewing the involved partitions with the intended semantics as the observer's inferences about the abstract information states from values. First, if the observer considered that the value $v$ resulted from evaluating $\op$ with values $w_1$, \ldots , $w_m$ for its operands, then it would infer from each value $w_i$ that the abstract information states is in $\block_{w_i}$ and thus in the intersection of these sets. Second, the observer normally cannot identify the actual operand values, but only a set of candidates for them. 
For each such candidate $(w_1, \ldots , w_m)$ the observer considers the possibility that the abstract information state is in the respective intersection for $(w_1, \ldots , w_m)$  and so concludes that the abstract information state is in the union of all these intersections.
Lastly, the branch $\symBranch$  and join $\symJoin$ of paths are interpreted as  follows: 
\begin{align}
\notag(\block_w^1)_{w \in \genRange_1} \symBranch  (\block_w^2)_{w \in \genRange_2}:=  &  \{ \block_v  \mid  \true \in \genRange_1, v \in \genRange_2,\block_v \neq \emptyset, \block_v = \block^1_\true \cap \block^2_v  \},\\
 (\block_w^1)_{w \in \genRange_1} \symJoin  (\block_w^2)_{w \in \genRange_2} := &
  \{ \block_v \mid \block_v \neq \emptyset,\, \block_v = \underset{v  \in \genRange_1}{\bigcup }\block^1_v \, \cup \, \underset{ v  \in \genRange_2}{\bigcup }\block^2_v \}.
\end{align}
The reader may find an example of the FlowTracker's interpretation of composite symbolic expressions in Example~\ref{Exa:TempView} of Section~\ref{Sec:Overview}.

Formally, we can show correct and non-interfering tracking  if we assume that the temporary views for basic information reactions are initialized correctly and define a symbolic interpretation function\\$\evalsymb: SymExpr \times  (\symbols \rightarrow Bool \cup Op)$\\\hspace*{2,4cm} $\times (\hvars \rightarrow (\genRange \pfn \pset{\bstates}) \times (\lvars \rightarrow \vals) \rightarrow$\\\hspace*{8cm}$(\hvars \rightarrow (\genRange \pfn \pset{\bstates}))$\\inductively in the way sketched above.
\begin{assumption}\label{Asmp:InitView}
For all $\bstate \in \bstates$, $\basicreq_{para} \in \basicreqs$, $v,w \in \vals$ it holds\\ $\evalprog (\basicreq_{para}) (v,\bstate ) = w $ iff $\bstate \in \block_w$ in $\initpartition (\basicreq_{para}, v)$.
\end{assumption}
\begin{theorem}[Flow tracking]\label{Thm:FlowTrack}
Let $\system$ be as specified in Table~\ref{Tab:Runs} and Table~\ref{Tab:RunsObs} satisfying Property~\ref{Def:Gradual} (gradual release),  and Property~\ref{Def:NoReadUp} (No-read-up) and Property~\ref{Def:NoWriteDown} (No-write-down), given in the appendix.
Let the FlowTracker of Table~\ref{Tab:Runs} translate the code $\cm{h}\code$ by $\transsymb(\transet(\cm{h}\code),\env)$ to $(\symb , \syminit )$ and interpret  each  symbolic expression $\symb(x)$ for $x\in\hvars$ by the function $\evalsymb$ as a change of the temporary view $\partitions (x)$.
Then, the system $\system$ satisfies Property~\ref{Def:Correctness} (Correctness) and Property~\ref{Def:NonInterference} (Non-interference) of flow tracking.
\end{theorem}
 
 \section{Further challenges and solutions for the Java-based instantiation}\label{Sec:FurtherChallenges}
 In Section~\ref{Sub:Impl} we have outlined our Java-based instantiation of the mediator framework with two focuses:
 first, on the initialization of the mediator in the preparatory phase of Figure~\ref{Fig:Scenario} and,
 second, in the working phase of Figure~\ref{Fig:Scenario},  on the synchronization of the components Program, FlowTracker and CIECensor through a Mediator class.
 The full functionality of the Java-based instantiation of the complete mediator framework may be summarized as follows:
  \begin{enumerate}
  \item template-based compilation of the mediator program to initialize the mediator framework according to the pattern in Section~\ref{Subsec:Mediator} such that the dynamic monitor, which consists of the FlowTracker and the CIECensor, is isolated from the program by the Mediator class as a proxy by means of Java-facilitated encapsulation;
 \item\label{Insta:InitView}initialization of temporary views according to Assumption~\ref{Asmp:InitView} for rows of relational data which each represents an abstract information state like in Example~\ref{Exa:AIS};
 \item flow tracking of protected realm fragments synchronized with program execution;
 \item\label{Insta:Generalization} policy-enforcement at declassification by means of subtree generalization schemes.
  \end{enumerate}
  This functionality is yet limited to mediator programs without loops and with finite enumeration types only. 
  Moreover, we eased the fully automated compilation of the mediator program
  by restricting the form of the program even more, for example to the use of object fields instead of program 
  variables. 
  However, the instantiation under these limitations still poses intricate challenges.
  In the following,  we point out several of such challenges related to the second and forth functionality and present preliminary solutions by way of example. 
  Regarding challenges related to other functionality in the list we refer the reader to Section~\ref{Sub:Impl}.

  \begin{figure}
  \includegraphics[width=.4\textwidth]{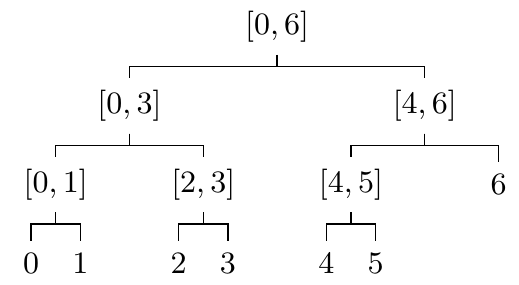}
  \caption{Generalization hierarchy for the finite integer domain $\{0,1,\ldots ,6\}$}\label{Fig:GenInt}
  \end{figure}
  \begin{example}
  Within the scenario described in Example~\ref{Exa:Scenario}, we focus on a different data source containing a row of an individual identified by the primary key $\ID$ which has two functionally dependent further attributes $D$ and $E$ each with a finite domain of integers $\mathit{dom}(D) = \mathit{dom}(E) = \{0,1,2,3\}$.  
  This source may be accessed by select and project queries similarly as the source considered in Example~\ref{Exa:Scenario}. 
  A generalization hierarchy may be defined such that it at least contains the domains of attributes $D$ and $E$ as shown in Figure~\ref{Fig:GenInt}.
  
  A first challenge is to overload the addition of integers with an operator $\oplus$ for the addition of intervals as generalized integers.
  The result of this operator must lay within the predefined, finite generalization hierarchy. 
  Hence, we set $[x_1,x_2] \oplus [y_1,y_2] = [x_1+y_1,x_2 + y_2]$ if the resulting interval is in the hierarchy. 
  Here, an integer $x$ is interpreted as the interval $[x,x]$.
  Otherwise, we take the minimal interval $[z_1,z_2]$ in the hierarchy (from bottom to top) such that $[x_1+y_1,x_2 + y_2]  \cap [0,6] \subset [z_1,z_2]$.
  For example, we may compute $[0,1] \oplus [0,1] = [0,3]$ and $[2,3] \oplus 1 = [0,6]$.
  Finally, if such a minimum does not exists as for $[2,3] \oplus [4,6]$ we define $[0,6]$ as the result.
  
  Now, we consider that the mediator should compute the sum of the columns $D$ and $E$ in the individual's row.
  For this simple computation, we may write the following two programs in pseudo-code 
  (not in Java to shorten notation) which use declassification assignments at different points throughout computation.
  \begin{multicols}{2}
  \vspace*{-1cm}
    \begin{small} 
\begin{gather*}
\intprocprog_1: x_{rea}\\
\text{1: }  x_1 := \basicreq (\text{project} , \{D\}))\\
\text{2: }  \icontrol{x_1}{l_{x_1}}\\
\text{3: }  x_2 := \basicreq (\text{project} , \{E\}))\\
\text{4: }  \icontrol{x_2}{l_{x_2}}\\
\text{5: }  x_{rea} := l_{x_1} \oplus l_{x_2}
  \end{gather*}
  \end{small}
\columnbreak
  \begin{small} 
\begin{gather*}
\intprocprog_2: x_{rea}\\
\text{1: }  x_1 := \basicreq (\text{project} , \{D\}))\\
\text{2: }  x_2 := \basicreq (\text{project} , \{E\}))\\
\text{3: }  x_3 := x_1 \oplus x_2\\
\text{4: }  \icontrol{x_3}{x_{reas}}
  \end{gather*}
  \end{small}
  \end{multicols}
  A second challenge now is to initialize the temporary views for the FlowTracker in lines~1 and~3 of program $\intprocprog_1$ and in lines~1 and~2 of program $\intprocprog_2$.
  To this end, the complete set of possible rows may be determined as $\bstates = \{ (id,D,E) \mid D,E \in \{0,1,2,3\} \}$.
  A block for the partition that results from the projection to $D$, for example, may be determined from its index $v$ by the 
  select query with selection predicate $D=v$ evaluated on $\bstates$ when $\bstates$ is stored as a relational database instance.
  This way the preimage of the result $v$ of the projection to $D$ is determined. 
  If the mediator accessed several rows, each related to another individual, the result of the projection might be a set $S$ of values the preimage of which is the union of all preimages each determined for a respective value $v \in S$.
  In general, the challenge of temporary view initialization is related to the problem of query inversion~\cite{BiPaSchw04}. 
  
  With the program $\intprocprog_1$, the FlowTracker computes the temporary view of $x_1$ and that of $x_2$ just by initializing them in the
  illustrated way, whereas with the program $\intprocprog_2$ the FlowTracker additionally computes the temporary view of $x_3$.
  The latter temporary view has seven blocks, one for each result value of adding integers between $0$ and $3$. 
  
  The additional computational effort of the FlowTracker during the execution of $\intprocprog_2$ is compensated by a possibly more informative result value for $x_{rea}$  in comparison to program $\intprocprog_1$. 
  To illustrate the decreased loss of information, we consider a confidentiality policy with two elements $\secret_{D=3}$ and $\secret_{E=3}$, 
  declaring that the value $3$ of $D$ and of $E$, respectively, is confidential.
  Based on this policy, the CIECensor determines the following security configurations:  $\secconf_1 = (\genRange_1' , \{\{3\}\})$ with $\genRange_1' = \{0,1,2,3\}$ for the declassification in line~1 and also in line~3 of program $\intprocprog_1$,  and $\secconf_2 = (\genRange_2' , \{6\})$ with $\genRange_2' = \{0,1,\ldots , 6\}$ for the declassification in line~4 of program $\intprocprog_2$.
  
  For each declassification, the CIECensor computes a generalized value with Algorithm~\ref{Algo:General} for subtree generalization.
  To follow each of the CIECensor's steps we take $\bstate=(id,2,1)$ for the row accessed by the program in the source data.
  First, in line~2 of program $\intprocprog_1$  the CIECensor selects the subtree with root $[2,3]$ for $\secconf_1$
  so that the value $2$ of $x_1$ is generalized to the interval $[2,3]$.
  Second, in line~3 the CIECensor again selects the same subtree, but here returns  the unmodified value $1$ of $x_2$ since that value is not an element of the selected subtree.
  Finally, the result $x_{rea}$ has the value $[2,3] \oplus 1 = [0,6]$ which is the smallest interval in the hierarchy that contains $[3,4]$.
  This loss of information in the result sent to the partner does not occur when computing the sum with program $\intprocprog_2$.
  In line~4 of program $\intprocprog_2$, the CIECensor select the subtree with root $[4,6]$ for $\secconf_2$, so that it returns the value $3$ of $x_3$,
  the sum of attributes $D$ and $E$, unmodified.
  
  The challenge with which mainly the program developer is burdened in our proposed framework is to balance the trade-off between computational effort and most informative reactions to the partner by inserting declassification assignments into the program code appropriately.
  An appropriate way for the developer to do so is to structure the mediator program in a modular way,  for example, to compute and declassify intermediate results for the reaction to the partner in dedicated methods.
  \end{example}

  \section{Conclusion}\label{Sec:RelatedWork}
   In this article, we have set a highly ambitious goal, namely how to provably enforce confidentiality of information by confining the information content of message data generated by program execution based on possibly diverse data sources.
    Towards this goal, we presented a mediator framework for a unified, holistic, history-aware control and applied a modular verification method to verify enforcement of a confidentiality policy on the unified view of the data sources, called abstract information state. 
  The control bases on or adapts existing technologies for (1) the isolation of processing realms, overcome solely by declassification;
  (2) tracking of explicit and implicit information flow from the abstract information state to program variables; (3) informative value generalization based on a generalization hierarchy, and (4) a Java-based instantiation of the framework.  
  
  In this section, we contrast our achievements for these four tasks with those in related work, and give hopefully constructive  ideas how the 
   achievements of the related fields could contribute to further improvements of our proposed framework.
 We are convinced that for feasible and applicable instantiations of the framework, in the long-term,  the expertise from different fields of research is needed.
 This conviction led us to spending much effort to find and exploit established methods for solutions of the four tasks.
 Research fields supporting our efforts  include information system integration and mediation~\cite{DoHaIv12}, data-program interfacing~\cite{Fo03}, logic-based inference control~\cite{Bi12a} and language-based information flow control with declassification~\cite{SaSa09} as further discussed in~\cite{Biskup2015Constructing}.

 Research on inference-usability confinement, e.g.~\cite{Bi12a,Bi13,BiBoGaSa14}, studies  how to design a control mechanism for a logic-based information system to confine the information content of a reaction to 
a  basic information request of diverse kinds,
such that the observer of communicated reaction data is always sure about the possibility that a confidential piece of information does not hold in the underlying information system.  
As a more general and comprehensive approach, in this article we consider program-based processing of requests so that our proposed control mechanism should comprise methods of both language-based information flow control and inference-usability confinement. 
 With our focus on such a unified control, we used an abstract representation of information like in~\cite{BiBoGaSa14} for the purpose of flow tracking in task (2), but for future instantiations of our framework we could use a logic-based representation of the temporary views and the observer's previous view.
 %, so that the determination of a security configuration would involve logical entailment instead of set containment.  
 In this case, the confidentiality policy could be stated more conveniently as  a set of logical sentences and Asumption~\ref{Asmp:InitView} can be established by results on view computations for particular basic information requests,  including queries to relational databases or incomplete databases, updates, see the summaries~\cite{Bi12a,Bi13} and revisions~\cite{BiTa15}. Such requests can be jointly offered by information mediation, to be integrated into our framework as outlined in~\cite{Biskup2015Constructing}.
 
Value generalization, which is applied in task~(3), is a fundamental method in database security to confine the information content of published data in a policy-compliant way, for example for \textit{k}-anonymity, see the survey~\cite{FuWaChYu10}.  
The extensive research on this method treats other important aspects not touched in this article, but relevant for the improvement of the framework in future work. 
One aspect is  the combination of generalization hierarchies of several  domains and
the efficient exploration of all candidates for optimal generalization schemes~\cite{LeDeRa05}. 
Another aspect is the minimality of generalization schemes under other cost measures than the height of the subtree, which is used for the scheme in Section~\ref{Sec:DistTab}. 
The work~\cite{BaAg05} presents an efficient strategy by which the candidate space is explored for minimizing any cost measure.
To prune the search space, cost lower-bounds are used which may be calculated by a generic method as a further contribution of the cited work.
For the proposed framework, such methods might be useful to estimate the loss of information 
caused by  declassification,  and to provide the program developer with this estimate to assist him in the proper placement of declassification assignments.
 
 Gradual release~\cite{AsSa07}, which in our framework guarantees the isolation of realms to be bypassed only via declassification for task (1), is  extended to the conditioned gradual release property in~\cite{BaNaRo08} which additionally confines the declassified information content.
 The  specification, which expresses such a property, combines a security level typing for gradual release with declarations of sentences in relational logic for declassification, which define the values of which expressions may flow to which low variables via which declassification procedures.
 The static program analysis employs a security type system for  the guarantee of gradual release and relational verification of the declassification procedures, linking the two results, typeability and verifiability, together by an analysis of execution paths via correspondence relations for the proof of conditioned gradual release. 
 
Our combination of gradual release via declassification with a dynamic monitor follows this idea, but replaces static verification in relational logic with dynamic mechanisms of the monitor, i.e., flow tracking and value generalization.
These mechanisms cannot be replaced by static relational verification for a policy-compliant declassification generally, because confidentiality policies often cannot be checked statically due to data-dependence.   
While we achieve data-level policy enforcement by means of value generalization here,  future work might investigate how by purely static program analysis schema-level policies might be enforced.
This has already been done for a security labeling of columns of a data table in~\cite{SchoepeEtAl2014SeLINQ} and might reasonably be extended to a declaration of confidential associations between columns.

Similarly to the task of flow tracking (2), a dynamic monitor achieves enforcement of assertions in relational logic by a dynamic tracking of valid assertions in~\cite{Chudnov2014RelationalLogic}. 
 The tracking considers alternative execution paths, which correspond to the actual execution path in a formal sense similar to correspondence relations in~\cite{BaNaRo08}.
However, like the FlowTracker, the monitor does not explicitly inspect all such paths, but tracks them implicitly as the set of all paths which agree with the actual path in the evaluation of particular expressions as specified in relational logic sentences.
If the monitor is unable to ensure the validity of assertions declared in the policy, it raises a security error,  a reaction less meaningful than value generalization.
However, the approximation of the set of corresponding execution paths might be useful to optimize the dynamic monitor in our framework. 

Another method, which might be adapted for approximating the observer's knowledge gain during flow tracking for task (4), is studied in~\cite{MaMaHiSri13}. 
Knowledge about a hidden part of memory is represented as a distribution which maps each possible hidden part to a decimal indicating the part's probability. 
Such knowledge is approximated by a polyhedron which defines probability bounds for all possible hidden parts of memory. 
Operators for knowledge update, which is essential to any method of inference control, track the effects of observable program execution on this knowledge.
The control enforces that the probability of each possible hidden part lays below a predefined threshold by updating the attacker's initial knowledge 
with observations from executing the complete program and rejecting the program if the threshold was exceeded. 

Also we are aware that a security type system approximately decides whether a program satisfies the gradual release property for task (1), and non-interference as its special case, with false negatives, and dynamic monitoring of the whole program execution may reduce false negatives~\cite{RuSa10}. 
However,  we chose security type systems, that guarantee the programming-language independent properties of gradual release and correspondence relations, as a unified approach. A type system can be enriched with other features such as flow lock policies~\cite{BrDeSa13} which define an event-based declassification mechanism. 

Section~\ref{Sub:Impl} highlighted two particular challenges of task~(4),  the Java-based instantiation of the mediator framework,  first, to synchronize flow tracking with low value computations by the program and, second, to bind temporary views to variables.  
The Symbolic PathFinder~\cite{PaViBuGeMeRu13} extends the Java Virtual Machine of the Java PathFinder with mechanisms which interweave symbolic execution with Java byte code execution and which as such relate to these two challenges. 
First,  listeners may be activated at particular events during byte code execution,  such as the transition from the open to the protected processing realm.
Furthermore, to attach symbolic information to variables, for each variable the object heap and the method stack store additional attributes which may be written and read by any byte code instruction given access to the variable. 
However, access to a method stack by another method is not allowed by the Java PathFinder Virtual Machine
which poses a problem to flow tracking as discussed in Section~\ref{Sub:Impl}. 
Finally,  several execution paths may be traversed in parallel according to the selection of paths by a choice generator.
However, the join of paths at certain events such as transition from the protected to the open realm is not supported by the Symbolic PathFinder.
Altogether, the Symbolic PathFinder does not help to address the two managed challenges of flow tracking.
 
 In conclusion, this work presents an extendible framework for the unified, holistic control
 of information flow from an abstract information state to a cooperation partner to effectively enforce 
 a partner-specific confidentiality policy,
 and exemplarily lays out a Java-based instantiation of this framework outlining first solutions
 for the challenges incurred in such a task. 
 
 \bibliography{literature}

\begin{thebibliography}{10}

\bibitem{ParagonWebsite}
Paragon website.
\newblock \url{http://www.cse.chalmers.se/research/group/paragon}.
\newblock Accessed: September 2016.

\bibitem{AsSa07}
A.~Askarov and A.~Sabelfeld.
\newblock Gradual release: Unifying declassification, encryption and key
  release policies.
\newblock In {\em 2007 {IEEE} Symposium on Security and Privacy (S{\&}P 2007)},
  pages 207--221. {IEEE} Computer Society, 2007.

\bibitem{BaDaGu12}
M.~Balliu, M.~Dam, and G.~L. Guernic.
\newblock Encover: Symbolic exploration for information flow security.
\newblock In S.~Chong, editor, {\em {IEEE} Computer Security Foundations
  Symposium -- {CSF} 2012}, pages 30--44. {IEEE} Computer Society, Los
  Alamitos, 2012.

\bibitem{BaNaRo08}
A.~Banerjee, D.~A. Naumann, and S.~Rosenberg.
\newblock Expressive declassification policies and modular static enforcement.
\newblock In {\em {IEEE} Symposium on Security and Privacy -- S \& P 2008},
  pages 339--353. {IEEE} Computer Society, Los Alamitos, 2008.

\bibitem{Bi12a}
J.~Biskup.
\newblock Inference-usability confinement by maintaining inference-proof views
  of an information system.
\newblock {\em International {J}ournal of {C}omputational {S}cience and
  {E}ngineering}, 7(1):17--37, 2012.

\bibitem{Bi13}
J.~Biskup.
\newblock Logic-oriented confidentiality policies for controlled interaction
  execution.
\newblock In A.~Madaan, S.~Kikuchi, and S.~Bhalla, editors, {\em Databases in
  Networked Information Systems -- {DNIS} 2013}, volume 7813 of {\em LNCS},
  pages 1--22. Springer, Berlin Heidelberg, 2013.

\bibitem{BiBoGaSa14}
J.~Biskup, P.~A. Bonatti, C.~Galdi, and L.~Sauro.
\newblock Optimality and complexity of inference-proof data filtering and
  {CQE}.
\newblock In M.~Kutylowski and J.~Vaidya, editors, {\em Computer Security --
  {ESORICS} 2014}, volume 8713 of {\em LNCS}, pages 165--181. Springer
  International Publishing, 2014.

\bibitem{BiPaSchw04}
J.~Biskup, J.~Paredaens, T.~Schwentick, and J.~V. den Bussche.
\newblock Solving equations in the relational algebra.
\newblock {\em {SIAM} J. Comput.}, 33(5):1052--1066, 2004.

\bibitem{Biskup2015Constructing}
J.~Biskup and C.~Tadros.
\newblock Constructing inference-proof belief mediators.
\newblock In P.~Samarati, editor, {\em Data and Applications Security and
  Privacy {XXIX} - 29th Annual {IFIP} {WG} 11.3 Working Conference ({DBSec}
  2015)}, volume 9149 of {\em LNCS}, pages 188--203. Springer, 2015.

\bibitem{BiTa15}
J.~Biskup and C.~Tadros.
\newblock Preserving confidentiality while reacting on iterated queries and
  belief revisions.
\newblock {\em Ann. Math. Artif. Intell.}, 73(1-2):75--123, 2015.

\bibitem{BiWe08}
J.~Biskup and T.~Weibert.
\newblock Keeping secrets in incomplete databases.
\newblock {\em Int. {J}. {I}nf. {S}ec.}, 7(3):199--217, 2008.

\bibitem{BrDeSa13}
N.~Broberg, B.~van Delft, and D.~Sands.
\newblock Paragon for practical programming with information-flow control.
\newblock In C.~Shan, editor, {\em Programming Languages and Systems -- {APLAS}
  2013}, volume 8301 of {\em LNCS}, pages 217--232. Springer International
  Publishing, 2013.

\bibitem{Bucur2011ParallelSymbolic}
S.~Bucur, V.~Ureche, C.~Zamfir, and G.~Candea.
\newblock Parallel symbolic execution for automated real-world software
  testing.
\newblock In C.~M. Kirsch and G.~Heiser, editors, {\em Proceedings of the Sixth
  European Conference on Computer Systems ({EuroSys} 2011)}, pages 183--198.
  {ACM}, 2011.

\bibitem{Chudnov2014RelationalLogic}
A.~Chudnov, G.~Kuan, and D.~A. Naumann.
\newblock Information flow monitoring as abstract interpretation for relational
  logic.
\newblock In {\em {IEEE} 27th Computer Security Foundations Symposium ({CSF}
  2014)}, pages 48--62, 2014.

\bibitem{DoHaIv12}
A.~Doan, A.~Y. Halevy, and Z.~G. Ives.
\newblock {\em Principles of Data Integration}.
\newblock Morgan Kaufmann, San Francisco, 2012.

\bibitem{Fo03}
M.~Fowler.
\newblock {\em Patterns of Enterprise Application Architecture}.
\newblock Pearson, Boston, 2003.

\bibitem{FuWaChYu10}
B.~C.~M. Fung, K.~Wang, R.~Chen, and P.~S. Yu.
\newblock Privacy-preserving data publishing: A survey of recent developments.
\newblock {\em ACM Comput. Surv.}, 42(4), 2010.

\bibitem{BaAg05}
R.~J.~B. Jr. and R.~Agrawal.
\newblock Data privacy through optimal k-anonymization.
\newblock In {\em Proceedings of the 21st International Conference on Data
  Engineering, {ICDE} 2005, 5-8 April 2005, Tokyo, Japan}, pages 217--228,
  2005.

\bibitem{LeDeRa05}
K.~LeFevre, D.~J. DeWitt, and R.~Ramakrishnan.
\newblock Incognito: Efficient full-domain k-anonymity.
\newblock In {\em Proceedings of the {ACM} {SIGMOD} International Conference on
  Management of Data}, pages 49--60, 2005.

\bibitem{MaMaHiSri13}
P.~Mardziel, S.~Magill, M.~Hicks, and M.~Srivatsa.
\newblock Dynamic enforcement of knowledge-based security policies using
  probabilistic abstract interpretation.
\newblock {\em Journal of Computer Security}, 21(4):463--532, 2013.

\bibitem{Nielson1999Principles}
F.~Nielson, H.~R. Nielson, and C.~Hankin.
\newblock {\em Principles of {P}rogram {A}nalysis}.
\newblock Springer, 1999.

\bibitem{PaViBuGeMeRu13}
C.~S. Pasareanu, W.~Visser, D.~H. Bushnell, J.~Geldenhuys, P.~C. Mehlitz, and
  N.~Rungta.
\newblock Symbolic pathfinder: integrating symbolic execution with model
  checking for java bytecode analysis.
\newblock {\em Autom. Softw. Eng.}, 20(3):391--425, 2013.

\bibitem{Phan2014AbstractModel}
Q.~Phan and P.~Malacaria.
\newblock Abstract model counting: a novel approach for quantification of
  information leaks.
\newblock In S.~Moriai, T.~Jaeger, and K.~Sakurai, editors, {\em 9th {ACM}
  Symposium on Information, Computer and Communications Security, ({ASIA} {CCS}
  2014)}, pages 283--292. {ACM}, 2014.

\bibitem{RuSa10}
A.~Russo and A.~Sabelfeld.
\newblock Dynamic vs. static flow-sensitive security analysis.
\newblock In {\em {IEEE} Computer Security Foundations Symposium -- {CSF}
  2010}, pages 186--199. {IEEE} Computer Society, 2010.

\bibitem{SaSa09}
A.~Sabelfeld and D.~Sands.
\newblock Declassification: Dimensions and principles.
\newblock {\em Journal of Computer Security}, 17(5):517--548, 2009.

\bibitem{SchoepeEtAl2014SeLINQ}
D.~Schoepe, D.~Hedin, and A.~Sabelfeld.
\newblock {SeLINQ}: tracking information across application-database
  boundaries.
\newblock In J.~Jeuring and M.~M.~T. Chakravarty, editors, {\em Proceedings of
  the 19th {ACM} {SIGPLAN} international conference on Functional programming,
  Gothenburg, Sweden, September 1-3, 2014}, pages 25--38. {ACM}, 2014.

\bibitem{Za16}
J.~Zarouali.
\newblock {Entwurf und Implementierung eines Rahmenwerks zur Inferenzkontrolle
  für Java-Applikationen mittels Paragon}.
\newblock Masterarbeit, TU Dortmund, 2016.

\end{thebibliography}
\newpage

%%%%%%%%%%%%%%%%%%%%%%%%%%%%%%%%%%%% Joachim %%%%%%%%%%%%%%%%%%%%%%%%%%%%%%%%%%%%%%%%%%%%%%%%%%%%%%%%%%%%
%%%%%%%%%%%%%%%%%%%%%%%%%%%%%%%%%%%%%%%%%%%%%%%%%%%%%%%%%%%%%%%%%%%%%%%%%%%%%%%%%%%%%%%%%%%%%%%%%%%%%%%%

\begin{appendix}
\setcounter{property}{4}
 \section{Correspondence relations as a proof tool}\label{Sec:Correspondence}
To argue about the observer's knowledge gain after declassification,  in particular for the proof of Theorem~\ref{Thm:Declassification},
we need to match declassification assignments in two runs $r$ and $r'$ that the observer regards as alternatives at times $t$ and $t'$, respectively, 
due to $\obsst (r,t) = \obsst (r',t')$.
 With such a motivation, we 
%review 
reuse correspondence relations and useful results from~\cite{BaNaRo08} in our notations.
As considered for Property~\ref{Def:Gradual} (Gradual release), an \textit{active} command is the command  evaluated next in the one-step semantics.
Moreover,  in the situation that $\code (r,t) \hasform \cm{h}\code \cm{; l}\code$ with the longest high-level prefix $\cm{h}\code$ of $\code (r,t)$ and a possibly empty low-level subprogram $\cm{l}\code$  
we call  $\cm{l}\code$  the \textit{L-continuation} of $\code (r,t)$ and write $\mathit{Lcont}(\code (r ,t ) )$.
This is exactly the situation in which flow tracking might start for $\cm{h}\code$ and end with the start of $\cm{l}\code$ according to case (1) of Table~\ref{Tab:Runs}.
As such, this situation is considered for inductive arguments in proofs concerned with flow tracking.
%A correspondence relation is a tool to match such situations in a pair of indistinguishable runs.
\begin{definition}[Correspondence (Definition 7.2 in~\cite{BaNaRo08})]\label{Def:CorrespRel}
Let $r$ and $r'$ be runs and $t$, $t'$ times.
A correspondence from $r$ to $r'$ [until $t$ and $t'$ respectively] is a relation $\Q{}{} \subseteq \{0, \ldots t\} \times \{0, \ldots ,t' \}$  
such that $\Q{0}{0}$ and for all $i,j$ with $\Q{i}{j}$ the following properties hold:
\begin{enumerate}
\item (state agreement)  $\mem_\low (r,i) = \mem_\low (r',j)$;
\item (level agreement)  active command of $\code (r,i)$  is low iff  active command of $\code (r',j)$ is low;
\item (code agreement $\low$)  $\code (r,i) = \code (r',j)$ if the active command is low;
\item  (code agreement $\high$)  $\mathit{Lcont}(\code (r,i) ) = \mathit{Lcont}(\code (r',j))$ if the active command is high;
\item (monotonicity) if $\Q{i}{j}$, $i < i'$, and $\Q{i'}{j'}$ then $j\le j'$; and symmetrically: if $\Q{i}{j}$, $j<j'$, and $\Q{i'}{j'}$ then $i \le i'$;
\item (completeness) for every $i \in \{0,\ldots , t\}$ there is some $j$ with $\Q {i}{j}$, and symmetrically. 
\end{enumerate}
\end{definition}
The following lemmata 
are ensured by every security type system that adheres to a weakened form of a no-read-up rule, as paraphrased  in Section~\ref{Sec:Overview} under  Item~1 (Isolation),  and a complementary no-write-down rule which both are defined after the lemmata.
As we have just anticipated as our main motivation for the use of correspondence relations, these lemmata ease inductive proofs of what knowledge the mediator subsystem reveals through declassification assignments.
First, Lemma~\ref{Lem:CorrespondenceEx}
enables us to match such assignments between the actual and an alternative run, that is a pair of runs at times $t$ and $t'$ such that $\lequiv{(r,t)}{(r',t')}$.
For such a corresponding pair of runs Lemma~\ref{Lem:CorrespondenceK} moreover enables us to relate the observer's perceptions of these runs and its knowledge thus gained. 

\begin{lemma}[Correspondence for $\lequiv{}{}$ (Lemma 7.3 in~\cite{BaNaRo08})]\label{Lem:CorrespondenceK}
Let $r$,$r'$ be runs and $t$, $t'$ times such that there exists a correspondence between $r$ and $r'$  until $t$ and $t'$, respectively.
Then, for the observer subsystem it holds $\lequiv{(r,t)}{(r',t')}$. 
\end{lemma}
\begin{lemma}[Correspondence (Lemma 7.5 in~\cite{BaNaRo08})]\label{Lem:CorrespondenceEx}
Let $r$, $r'$ be runs and $t$, $t'$ times such that $\lequiv{(r,t)}{(r',t')}$ holds for the observer subsystem and in run $r$ at time $t$ and in run $r'$ at time $t'$ the active commands are low.
Then there is a correspondence from $r$ to $r'$ until $t$ and $t'$, respectively.
\end{lemma}

If low computation steps depend only on what the observer is able to perceive, here low memory, its change and program termination (Table~\ref{Tab:RunsObs}),  such steps must be taken correspondingly in the actual and an alternative run during which the observer makes equal observations. 
The latter consequence is what Lemma~\ref{Lem:CorrespondenceEx} says, whereas the premise is subject of the following no-read-up rule, Property~\ref{Def:NoReadUp}.
This rule says that low computation steps, except for declassification assignments, on low memory are determined by low memory only.\footnote{
This is an intuitive, yet a little imprecise reading of the rule since the rule neglects the effect of the computation step on high memory.
High memory may be neglected because the lemmata concern the observer subsystem $\obsst$ and the existence of correspondence relations only. 
}
Putting it differently, such steps are executed as if high memory and abstract information state were not accessible at all.
\begin{property}[No-read-up without declassification (Lemma 6.3 b in~\cite{BaNaRo08})]\label{Def:NoReadUp}
Let the active command in a program $\code$ be low, but no declassification assignment, and be evaluated in the step $\evalprog ( \code , \mem , \bstate) =  (\code^{succ} , \mem^{succ}, \bstate)$ given any memory $\mem$ and  any abstract-information state $\bstate$.
Then, for every memory $\mem'$ with $\mem_\low = \mem_\low'$ there exists ${\mem'}^{succ}$ such that for every $\bstate' \in \bstates$ it holds  $\evalprog ( \code , \mem' , \bstate') =  (\code^{succ} , {\mem'}^{succ}, \bstate')$ and $\mem^{succ}_\low = {\mem'}^{succ}_\low$.
\end{property}

Single steps in a sequence of high level program execution in a run may be matched with high level steps in another run and another sequence by a correspondence relation.
Computations must not be the same in the matched steps  (code agreement $\high$), but so must the low memory states. 
This only works if high computation steps do not change low memory as stated by the following no-write-down rule.
\begin{property}[No-write-down (Lemma 6.3 c in~\cite{BaNaRo08})]\label{Def:NoWriteDown}
Let the active command in a program $\code$ be high and evaluated in the step $\evalprog ( \code , \mem , \bstate) =  (\code^{succ} , \mem^{succ}, \bstate)$ given any memory $\mem$ and abstract-information state $\bstate$.
Then,  it holds $\mem_\low = \mem^{succ}_\low$.
\end{property}
 
 Indeed the two properties suffice to establish the two lemmata.
 \begin{proposition}\label{Prop:Correspondence}
 Let $\intproclang$  be a programming language with a one-step semantics $\evalprog$. 
 Let $\env$ be a security level inference for $\intproclang$ which adheres to the rules in Property~\ref{Def:NoReadUp} and Property~\ref{Def:NoWriteDown}.
 Then, for the mediator subsystem of Table~\ref{Tab:Runs} and the observer subsystem $\lequiv{}{}$ of Table~\ref{Tab:RunsObs} Lemma~\ref{Lem:CorrespondenceK} and Lemma~\ref{Lem:CorrespondenceEx} hold.
 \end{proposition}
 
\section{Sketch of induction step for Theorem~\ref{Thm:Declassification}}
By induction on time $t$ we show the following relationships between the observer's knowledge $\K$, the observer's previous view $\view$ and
the observer's inferred abstract information set $\infintbel$, respectively:
\begin{align}
\label{Equ:PrevView}\K (r,t) & = \view (r,t)\\
%\intertext{and  if the preconditions of the theorem hold for $(r,t)$}
\label{Equ:Declass}\K (r,t+1) & = \K (r,t) \cap \infintbel (\partitions (r,t) (x_{src}) ,\disttab ,\secconf , g ).
\end{align}
%\noindent\emph{Base Case}. 
%It holds $\K(r,0) = \bstates = \view (r,0)$ for all runs $r$ by definition of $\K$ and $\system$.

%\noindent\emph{Inductive Case}.
% \begin{reasonbycases}
% \item  
\noindent \textit{Case} 1: Let run $r$ be  such that $\code (r,t) \hasform \icontrol{x_{src}}{x_{dest}};\cm{rest}\code$ where $x_{src}\in\hvars$ and $x_{dest} \in \lvars$.
% (preconditions of the theorem).
%\\
 In this case, \eqref{Equ:PrevView}  follows from~\eqref{Equ:Declass} given the computation of $\view$ by Algorithm~\ref{Algo:Censor}.
%  \begin{reasonbycases}
% \item
 
\noindent ``$\subseteq$ in~\eqref{Equ:Declass}'': Consider $\bstate \in \K (r,t+1)$ 
%for the proof of $\subseteq$ in~\eqref{Equ:Declass}.\\
%By definition of $\K$ there is  
and let a run $r'$ and a time $t'$ be such that 
\begin{equation}\label{Eq:Thm:Declass2}
 \bstate (r',0) = \bstate \text{ and } \obsst (r',t') = \obsst (r,t+1).
\end{equation}

Consider the latest times in both runs, in run $r$ until time $t+1$ and in run $r'$ until time $t'$,  when the active command  is low. 
According to Table~\ref{Tab:RunsObs} execution of an high active command cannot be observed, so that we can assume that times $t+1$ and $t'$ are such times.
In this situation, by Lemma~\ref{Lem:CorrespondenceEx},  due to the equality $\obsst (r',t') = \obsst (r,t+1)$ there is a correspondence $\Q{}{}$ from $r$ to $r'$ until $t+1$ and $t'$ respectively.

We exploit this correspondence to match the declassification assignment in both runs.
In the considered case, the active command in $(r,t)$ is  $\icontrol{x_{src}}{x_{dest}}$ which is always low level if $x_{dest}$ is as in this case.
Since the correspondence is complete and monotone, there is $t'' \le t'$ with $\Q{t}{t''}$.
Since $\Q {t} {t''}$,  we even know that $\code (r,t) = \code (r', t'')$ due to the level agreement and code agreement $\low$. 
Therefore, the active command in $(r',t'')$ is $\icontrol{x_{src}}{x_{dest}}$, too.

To prove $\subseteq$ in~\eqref{Equ:Declass}, we show first that run $r'$ at $t''$ clusters the same blocks of the same $\genRange$-indexed partition as $r$ does at $t$. 
Then we conclude by showing that the abstract information state of run $r'$, which is $\bstate$ by~\eqref{Eq:Thm:Declass2}, lies in this cluster.

By Lemma~\ref{Lem:CorrespondenceK}, the correspondence $\Q {t} {t''}$ between $r$ and $r'$ implies $\obsst (r,t) = \obsst (r',t'')$. 
Hence, by Property~\ref{Def:NonInterference} (Non-interference) the temporary views agree  $\partitions (r,t)(x_{src}) = \partitions (r',t'')(x_{src})$ for $x_{src}$ to be declassified in both $(r,t)$ and $(r',t'')$. Also do the previous views agree by the induction hypothesis for~\eqref{Equ:PrevView}.
Hence, during declassification, by Algorithm~\ref{Algo:Censor} the censor computes the same security configuration 
%of Definition~\ref {Def:SecConfig} 
and thus selects the same row in the distortion table.
Since the observations in both runs agree in the next time step $t+1$ and $t''+1$ (until $t'$), respectively, by~\eqref{Eq:Thm:Declass2},  in both $(r,t)$ and $(r',t'')$  the same generalized value $g$ is assigned to the destination variable $x_{dest}$.
This value determines the inferred abstract information $\infintbel$ of~\eqref{Equ:InfAbstInf}, given the temporary view of $x_{src}$ and the row in the distortion table, which thus is the same set of states, as a cluster of the original blocks in the temporary view of $x_{src}$, in both runs at times $t+1$ and $t''+1$, respectively. 

Finally, we show $\bstate \in \K (r,t) \cap \infintbel$. 
First, since knowledge is monotone from $\bstate \in \K (r, t+1)$ it follows $\bstate \in \K (r,t)$.
Second, the abstract information state $\bstate$, which is $\bstate(r',0)$ by the choice of $r'$ in~\eqref{Eq:Thm:Declass2}, is in  $\infintbel$ determined in $(r,t)$.
To show this,  we consider run $r'$ at time $t''$ in which the censor determines the same set $\infintbel$ as in $(r,t)$ by our previous arguments.
In run $r'$ at time $t''$, by Property~\ref{Def:Correctness} there is a block $\block_u$ in $\partitions (r',t'')(x_{src})$ with $\mem (r',t'')(x_{src})=u$ and $\bstate = \bstate(r',0) \in \block_u$.
Moreover, we know that value $u$ is generalized to the observed value $g$ (possibly $u=g$) so that $\block_u \subseteq \infintbel$.
%With these two arguments, we  finally proved $\bstate \in \K (r,t) \cap \infintbel$. 

%\item For the proof of 
\noindent ``$\supseteq$ in~\eqref{Equ:Declass}'': Consider $\bstate \in \K (r,t) \cap \infintbel$. 
We can proceed along a similar line of argumentation as 
%in the previous case
before, showing that the runs $r'$ and $r$ with $\obsst (r',t'') = \obsst (r,t)$ and $\bstate (r',0) = \bstate$ use the same row in
the distortion table.
To conclude, we claim that the same generalized value is transferred in both runs, which is observed at $t''+1$ and $t+1$, respectively.
Due to equal observations at those times it finally follows $\bstate = \bstate (r',0) \in \K (r, t+1)$.

We show that in both runs  the same generalized value is transferred to $x_{dest}$.
In $(r,t)$, from the cluster of blocks that is put into union to $\infintbel $, we take the block $\block_u$ that contains $\bstate (r',0)$. 
This is possible because by precondition it holds $\bstate (r',0) \in \infintbel$.
First, by Property~\ref{Def:NonInterference} (Non-interference) this block $\block_u$ is also a block in the partition $\partitions (r',t'')$.
Second, by taking Property~\ref{Def:Correctness} (Correctness) into account in each of the two runs, it follows that $\mem (r,t)(x_{src}) = u$ and $\mem (r',t'')(x_{src})=u$.
Recalling the arguments of the previous case, we know that the censor generalizes $u$ to the same value $g$ in both runs due to equal security configurations. 
%  \end{reasonbycases}
% \item 

\noindent \textit{Case} 2: Let $r$ be such that $\code (r,t) \hasform \icontrol{x_{src}}{ x_{dest}};\cm{rest}\code$ and $x_{src}\in\lvars$.
 Due to the monotonicity of knowledge, which ensures $\K (r,t+1) \subseteq \K (r,t)$, we only need to consider $\bstate \in \K (r,t)$ and show $\bstate \in \K (r,t+1)$. 
 As in the previous case, we consider a run $r'$ with $\bstate (r',0) = \bstate $ and $\obsst (r',t') = \obsst (r,t)$. As exemplified above, we use a correspondence relation to show that $\icontrol{x_{src}}{x_{dest}}$ is the active command in $(r',t')$, too, and, hence, due to the low label of $x_{src}$ observations in both runs at times $t+1$ and $t'+1$ are equal. 
 This equality shows $\bstate = \bstate (r',0) \in \K (r,t+1)$.
%\item  

\noindent \textit{Case} 3: Let $r$ be such that $\code (r,t) \hasform \icontrol{x_{src}}{ x_{dest}};\cm{rest}\code$ and $x_{dest}\in\hvars$.
We may proceed as in the previous case noting that because both runs write to the high variable $x_{dest}$ there are no observations at times $t+1$ and $t'+1$.

\noindent \textit{Case } 4: Let  $\icontrol{x_{src}}{x_{dest}}$ be not the active command in $(r,t)$.
%\\
 We use gradual release in Property~\ref{Def:Gradual} and the induction hypothesis to show~\eqref{Equ:PrevView},
 whereas \eqref{Equ:Declass} according to Theorem~\ref{Thm:Declassification} solely refers to the case where the active command is a declassification
 assignment with high source and low destination.\\[1ex]
% \end{reasonbycases}
%
\section{Sketch of induction step for Theorem~\ref{Thm:Conf}}
%\noindent \textbf{Sketch of Induction Step for Theorem~\ref{Thm:Conf}.}
%\noindent\emph{Base Case}.
%By definition of $\K$ and $\system$ it holds $\K (r,0) = \bstates$ for all runs $r$ so that Property~\ref{Def:Conf} holds by  Assumption~\ref{Asmp}.
%
%\noindent\emph{Inductive Case}.
If in Table~\ref{Tab:Runs} any case applies, but the value generalization by the CIECensor in case (3) of the table, we can argue that $\K (r,t) = \K (r,t+1)$ holds by reusing respective arguments from the proof sketch of Theorem~\ref{Thm:Declassification} and then use the induction hypothesis. 
Otherwise, the CIECensor determines a security configuration $\secconf = (\genRange' , \vioSets )$ for the declassification of $x_{src}$  with $x_{src}\in\hvars$, using the temporary view $(\block_w)_{w\in\genRange'}$ of $x_{src}$.
By Property~\ref{Def:Correctness} (Correctness) this view covers $\K (r,t)$ so that by~\eqref{Equ:PrevView} it holds $\K (r,t) \subseteq\view (r,t)  
%\cap \underset{w \in \genRange'}{\bigcup} \block_w $. 
\cap \bigcup_{w \in \genRange'} \block_w $.  
Therefore, the whole range of values $\genRange'$ cannot be in any of the violating sets $\vioSets$ contained in $\secconf$ 
%because $\K (r,t) \not\subseteq \secret$ for all $\secret \in \conf$ 
by induction hypothesis.

Finally, we assume the contrary that there exists $\secret \in \conf$ such that  $\K(r,t+1)\subseteq \secret$ which by Theorem~\ref{Thm:Declassification} can be rewritten to  $\view (r,t) \cap \infintbel \subseteq \secret$ in case (3) of Table~\ref{Tab:Runs}.
This assumption implies $\{u \in \genRange' \mid \disttab (\secconf , u ) = g \} \in \secconf$ since $\infintbel$ is defined as the union of blocks with  such an index $u$. 
Because we first argued that $\genRange' \not \in \secconf$, the latter conclusion contradicts Requirement~2 in Definition~\ref{Def:DistTab} of distortion tables. \\[1ex]
 \section{Sketch of induction step for Theorem~\ref{Thm:FlowTrack}}
%\noindent 	\textbf{Sketch of Induction Step for Theorem~\ref{Thm:FlowTrack}.}
%	
 We will show by induction on the times $t$ in which the FlowTracker's status is $\ftidle$ in run $r$, 
thus including all times of declassification according to case (3) of Table~\ref{Tab:Runs}, 
that for all $r' \in \system$ and $t'\in\mathbb{N}_0$ such that $\obsst (r',t') = \obsst (r,t)$ it holds:
\begin{gather}
\label{Equ:FTNonInterfereEquCodes}\code (r,t) = \code (r' ,t'),  \\
\label{Equ:FTNonInterfereEquPartitions} \partitions (r,t)  = \partitions (r',t') \text{ with domain } \hvars, \\
\label{Equ:FTCorrectness}\text{for all }x\in\hvars\;  \bstate (r,t) \in \block_v \text{ in } \partitions (r,t)(x) \text{ iff }\mem (r,t)(x) = v, \text{and}\\
\label{Equ:FTPartitions}\text{for all }x\in\hvars\; \partitions (r,t)(x)  \text{ forms a partition covering }\K (r,t).
\end{gather}
%\noindent\emph{Base Case}. 
%For $t=t'=0$ the commands $\code(r,0)$ and $\code (r',t')$ both are the whole interaction processing program and for each $x \in \hvars$ the temporary views are initialized to $(\bstates_w)_{w \in \{v\}}$  with %a constant default value $v$ in both runs.

%\noindent\emph{Inductive Case}.
The interesting case is where at time $t-1$ the FlowTracker's status is $\fttracking$ so that at time $t$ the FlowTracker changes the temporary views.
Let $t_{start}$ be the latest time before in run $r$ where the status is $\ftidle$.  
According to the definition of runs in Table~\ref{Tab:Runs}, the program is 
$\code (r, t_{start}) \hasform \cm{h}\code ;\cm{l}\code$, a sequence of  a high-level subprogram $\cm{h}\code$ and a low-level subprogram $\cm{l}\code$, 
so that for any other such sequence $\cm{h}\code';\cm{l}\code'$ the subprogram  $\cm{h}\code'$ is a prefix of $\cm{h}\code$.

Since $\obsst (r',t') = \obsst (r,t)$, 
%holds, 
there is a correspondence $\Q{t}{t'}$ between $r$ and $r'$ by Lemma~\ref{Lem:CorrespondenceEx}.
The program in run $r'$ at a corresponding time $t'_{start}$, determined by $\Q{}{}$, is the same  by the induction hypothesis, using that 
 $\obsst (r',t'_{start})=\obsst (r,t_{start})$ 
%holds 
by Lemma~\ref{Lem:CorrespondenceK}.  
For the same reason, all other arguments of the FlowTracker's computation
 $\eval(\transsymb(\transet(\cm{h}\code),\env))(\partitions , \mem_\low )$ in $(r,t_{start})$ and $(r',t'_{start})$ 
are equal (the low memory due to state agreement low).
%Altogether
So, we showed~\eqref{Equ:FTNonInterfereEquCodes} and the equality in~\eqref{Equ:FTNonInterfereEquPartitions}.

Regarding the domain in~\eqref{Equ:FTNonInterfereEquPartitions} and the equivalence of~\eqref{Equ:FTCorrectness}, we will consider the execution tree $T$ generated from $\cm{h}\code$ preserving the one-step semantics. 
Moreover, we will consider the memory $\mem_n$ produced by $T$ after processing a node $n$ and all expressions $expr$ defined over high variables and over expressions in the image of $\syminit$.
Let Algorithm~\ref{Algo:TransExpr} translate  $expr$ to  $symexpr$ in the symbolic state $\symb (n)$ with $\syminit$ and $\env$. 
Based on Assumption~\ref{Asmp:InitView} (Temporay view initialization) and Property~\ref{Def:Gradual} (Gradual release),  we can show by induction on the tree structure that for all nodes $n$ of $T$ 
 the domain of $\symb (n)$ is $\hvars$ for~\eqref{Equ:FTNonInterfereEquPartitions} and it holds $\evalprog (expr) (\mem_{n}, \bstate) = v $  iff $\bstate \in \block_v \in \evalsymb (symexpr)(\partitions , \mem_\low )$. 
%To do so, for the operators in $expr$ we can use the arguments which we illustrated in Sect.~\ref{Sec:FlowTracker} 
%for the symbolic interpretation of an operator $\op \in \Theta_m$. Also we use that the tree does not change low memory since $\cm{h}\code : \high$.
Since this way we can verify correctness for the translation of all relevant operations of the tree along a path, including branch conditions, we can show correctness of the translation of the complete tree for~\eqref{Equ:FTCorrectness}. 

 Finally,  we 
%need to 
verify that  for all $x \in \hvars$ $\partitions (r,t)(x)$ is indeed an $\genRange$-indexed partition covering $\K(r,t)$.
% and do so in two steps.
First, let  $\bstate \in \K (r,t)$. We can show that there is $\block_v$ in $\partitions (r,t)$ which contains $\bstate$ by selecting a run $r''$ 
such that $\bstate = \bstate (r'',0)$ and $\obsst (r'',t'') = \obsst (r,t)$ at a time $t''$,
then using~\eqref{Equ:FTCorrectness} to first find the the desired block in $\partitions (r',t'')(x)$ with the block index $v = \mem (r',t'') (x)$ and finally
concluding with~\eqref{Equ:FTNonInterfereEquPartitions} that the block found is also in $\partitions (r,t)(x)$.
Second, let $\block_v$ and $\block_{v'}$ be two blocks in $\partitions (r,t)(x)$ with $x \in \hvars$. 
We can show that $\block_v \cap \block_{v'} \neq \emptyset$ implies $v =v'$ and thus $\block_v = \block_{v'}$.

\section{Sketch of proof for Proposition~\ref{Prop:General}}
%\noindent \textbf{Sketch of Proof for Proposition~\ref{Prop:General}.}
%
First, we argue that Algorithm~\ref{Algo:General} indeed computes a function on $(\{\genRange' \in \pset{\genRange}\} \times \pset{\pset{\genRange'}\setminus \genRange'}) \times \genRange$.
So let $\secconf = (\genRange',\vioSets) \in \{\genRange' \in \pset{\genRange} \} \times \pset{\pset{\genRange'}\setminus \genRange'}$ and $w \in \genRange$. Then, we claim the following three points:
(1) an $\secconf$-subtree generalization scheme of Definition~\ref{Def:GenScheme} exists, and hence can be computed in line~\ref{Algo:DetermineScheme} by exhaustively searching the finite space of all candidates. 
(2) This scheme is uniquely determined by $\secconf$ and the generalization hierarchy $\taxtree$.
Moreover, (3) if existent, the subtree $\gtree$ in line~\ref{Algo:DetermineSubtree} is uniquely determined by $\gscheme$ and $w\in\genRange$.
So, following all three points, the algorithm computes a function.

We justify each of the above points:
(1) The set $\gscheme = \{ \genRange \}$ is always a candidate which satisfies all requirements of Definition~\ref{Def:GenScheme}, but minimality. 
Minimality is only fulfilled if there is no other candidate.  Regarding the third requirement, we note that for all $I \in \vioSets$ it holds $I \subseteq \genRange$ and $\genRange \cap \genRange' \not \subseteq I$ since $I\subseteq \genRange'$, but $I \neq \genRange'$ by Definition~\ref{Def:SecConfig} (Security Configuration).
(2) Let $\gscheme$ and $\gscheme'$ be two $\secconf$-subtree generalization schemes.  We show that for any $\gtree \in \gscheme$ there exists $\gtree' \in \gscheme'$ such that $\gtree'=\gtree$.  The proof of that will only exploit that $\gscheme$ is a subtree generalization scheme so that  we could show the same for $\gscheme'$. Hence, the two sets are equal.
Starting the proof, we know that $\gscheme \gle \gscheme'$ by the forth requirement which particularly says that there exists $\gtree' \in \gscheme'$ such that $\gtree \subseteq \gtree'$. Likewise, there exists $\gtree^\ast \in \gscheme$ such that $\gtree' \subseteq \gtree^\ast$ so that we obtain the chain $\gtree \subseteq \gtree' \subseteq \gtree^\ast$.
Since $\gtree, \gtree^\ast \in \gscheme$ and all such sets are disjoint, if unequal, it follows $\gtree = \gtree^\ast$ and hence $\gtree = \gtree'$.
(3) This point is obvious because all subtrees are pairwise disjoint.

Now, let $\disttab$ denote the function as computed by the algorithm.  We conclude by verifying the two properties of a distortion table.
(1) It is obvious that, if $\vioSets = \emptyset$, then $\gscheme = \emptyset$ is the $\secconf$-subtree generalization scheme and hence Algorithm~\ref{Algo:General} returns $w$ on every input value $w$.
(2) We show that for all $\secconf = (\genRange',\vioSets) \in \{\genRange' \in \pset{\genRange}\} \times \pset{\pset{\genRange'}\setminus\genRange'}$ and $g \in \genRange$ such that $\{ w \in \genRange \mid \disttab (\secconf , w) = g \} \cap \genRange' \neq \emptyset$ there does not exist $I\in\vioSets$ such that  $\{ w \in \genRange \mid \disttab (\secconf , w) = g \} \cap \genRange' \subseteq I$. 
We need to discuss two cases.
\begin{reasonbycases}
\item For all $\gtree$ in the $\secconf$-subtree generalization scheme $\gscheme$ it holds  $g \not\in\gtree$.\\
In this case, it holds that $\{ w \in \genRange \mid \disttab (\secconf , w ) = g \} = \{g\}$. Assume that there is $I \in \vioSets$ such that $\{g\}  \cap \genRange' \subseteq I$ and $\{g\} \cap \genRange' \neq \emptyset$. 
Hence, we obtain $g \in I$.  Since $\gscheme$ is a subtree-generalization scheme, it follows that there is a selection $S$ of subtrees  from $\gscheme$ such that $I \subseteq \underset{\gtree \in S}{\bigcup} \gtree$. This contradicts the case considered.
\item There is $\gtree$ in the $\secconf$-subtree generalization scheme $\gscheme$ such that $g\in\gtree$.\\
Then, by the definition of the algorithm this $\gtree$ has the root value $g$ and all values of $\gtree$ and only these are generalized to $g$, thus $\{ w \in \genRange \mid \disttab ( \secconf , w ) = g \}= \gtree$.
Assume that there exists $I \in \vioSets$ such that $\gtree \cap \genRange' \subseteq I$ and $\gtree \cap \genRange' \neq \emptyset$. By the definition of the generalization scheme there is a selection $S$ of subtrees  from $\gscheme$ such that $I \subseteq \underset{\gtree' \in S}{\bigcup} \gtree'$.
Putting it all together, we obtain that $\emptyset \neq \gtree \cap \genRange' \subseteq I  \subseteq \underset{\gtree' \in S}{\bigcup} \gtree'$ and hence $\underset{\gtree' \in S}{\bigcup} \gtree' = \gtree$, because all different sets are pairwise disjoint. 
This means that $\underset{\gtree' \in S}{\bigcup} \gtree' \cap \genRange' =  \gtree \cap \genRange' \subseteq I$, contradicting $\gtree' \cap \genRange' \not \subseteq I$ for all $\gtree' \in S$ as required when $S$ has been selected.
\end{reasonbycases}

\section{Sketch of proof for Proposition~\ref{Prop:Correspondence}}
\subsection{Claim: Lemma~\ref{Lem:CorrespondenceK} implied by Property~\ref{Def:NoReadUp} and Property~\ref{Def:NoWriteDown}}
%\noindent \textbf{Sketch of Proof for Proposition~\ref{Prop:Correspondence}.}
%
By induction on time $t$ we show that if there is a correspondence from $r$ to $r'$ until $t$ and $t'$ then it follows 
$\lequiv{(r,t)}{(r',t')}$. We will limit the proof sketch to the induction step.
By monotonicity and completeness of Definition~\ref{Def:CorrespRel}, it holds $\Q{t}{t'}$ and  $\Q{t-1}{t''}$ for some $t''\le t'$ and $\Q{t}{t'''}$ for all $t''' \in \{t''+1,\ldots , t'\}$.  
The induction hypothesis says that $\lequiv{(r,t-1)}{(r',t'')}$ holds. 
In the case of  $t''<t'$, we will argue that from the induction hypothesis and $\Q{t}{t''+1}$ the equality $\obsst ( r ,t ) = \obsst ( r', t''+1)$ follows.
This argument may be repeated for all $t''' \in \{t'' + 2 , \ldots , t'\}$ so that $\obsst (r,t) = \obsst (r', t''')$ also holds for all such $t'''$.
Finally, we consider the special case of $t''=t'$.

\begin{reasonbycases}
\item The active command in $(r,t-1)$ is low.\\
Code agreement $\low$ implies that the active command is the same in $(r',t'')$. 
If it is not an assignment or one to a high variable, then the observer cannot perceive its execution in both runs according to Table~\ref{Tab:RunsObs} so that it holds $\obsst(r,t) = \obsst (r, t-1)$ and $\obsst(r',t''+1) = \obsst (r' , t'')$.  By induction hypothesis it follows $\obsst (r,t) = \obsst (r',t''+1)$.

If the active command in $(r,t-1)$ assigns to a low variable, then by state agreement for $\Q{t}{t''+1}$ in $(r,t'')$ the same value is written and thus observed according to Table~\ref{Tab:RunsObs}. 
By induction hypothesis it follows $\obsst (r,t) = \obsst (r',t''+1)$.
\item The active command in $(r,t-1)$ is high.\\
By level agreement for $\Q{t-1}{t''}$ the active command in $(r',t'')$ is also high.
Property~\ref{Def:NoWriteDown} (No-write-down) ensures that low memory does not change in both runs so that there are no additional observations in both runs at times $t$ and $t''+1$, respectively, according to Table~\ref{Tab:RunsObs} and hence it holds $\obsst (r,t) = \obsst (r',t''+1)$ by induction hypothesis.
\item Other cases (of computation in $(r,t)$) according to Table~\ref{Tab:Runs}.\\
In the table, we see that computations by the mediator other than program execution, which is treated by the above cases, are the computation of
the function $\evalsymb$ in case (1) (Start flow tracking) of the table and  computation of the function $\censor$ in case (3) (Generalize value by CIECensor) of the table.
These computations do not change low memory, as can be seen from the table, and hence cannot be observed according to Table~\ref{Tab:RunsObs}.
\item Special case $t''=t'$.\\
By choice of $t''$,  it holds $\Q{t-1}{t''}$ and hence $\Q{t-1}{t'}$. The induction hypothesis, applied to the correspondence $\Q{t-1}{t'}$, says that $\obsst (r,t-1) = \obsst (r', t')$ holds. 
 Moreover, as mentioned above, it holds $\Q{t}{t'}$. 
Due to both correspondences from $t-1$ and $t$ to $t'$, the active command in $(r,t-1)$ cannot be low.
Otherwise, code agreement low for both $\Q{t-1}{t'}$ and $\Q{t}{t'}$ would imply  $\code (r,t-1)\hasform\code (r,t)$, but, before program termination, in Table~\ref{Tab:Runs} each transition in a run alters the current program.
Hence,  the active command in $(r,t)$ is high.

According to Property~\ref{Def:NoWriteDown} (No-write-down),   the high active command in $(r,t-1)$ may not change low memory, so that there are no additional observations in $(r,t)$ according to Table~\ref{Tab:RunsObs} and it holds $\obsst (r,t) = \obsst (r, t-1) = \obsst (r' , t')$.
\end{reasonbycases}

\subsection{Claim: Lemma~\ref{Lem:CorrespondenceEx} implied by Property~\ref{Def:NoReadUp} and Property~\ref{Def:NoWriteDown}}
By induction on time $t$ we show that if it holds $\lequiv{(r,t)}{(r',t')}$ and the active commands in $(r,t)$ and $(r',t')$ are low then there is a correspondence from $r$ to $r'$ until $t$ and $t'$. We will limit the proof sketch to the induction step.

Let $t_{pre}^\ast<t$ be the latest previous time where the active command is low in $r$, and let  ${t'}_{pre}^\ast <t'$ be the like time in $r'$.
The induction hypothesis says that there is a correspondence from $r$ to $r'$ until $t_{pre}^\ast$ and ${t'}^\ast_{pre}$.
First, we start with $t^\ast_{pre}+ 1$ and ${t'}^\ast_{pre}+1$.
\begin{reasonbycases}
\item  $t_{pre}^\ast +1 \neq t$  and ${t'}^\ast_{pre} + 1\neq t'$.\\
We extend the correspondence to $\Q{t_{pre}^\ast +1}{{t'}^\ast_{pre} + 1}$  and justify that for this extension the properties of a correspondence relation 
are still fulfilled.

First, by choice of $t^\ast_{pre}$ and ${t'}_{pre}^\ast$, the active commands in $(r,t_{pre}^\ast +1)$ and $(r',{t'}_{pre}^\ast +1)$ are high and
 moreover we claim that $\mathit{Lcont}( \code (r, t_{pre}^\ast + 1) )= \mathit{Lcont}( \code (r', {t'}_{pre}^\ast +1))$ holds.
Hence, level agreement and code agreement high are fulfilled.

As we justify in the following the latter equality of the longest low level suffixes follows from the correspondence from $r$ to $r'$ until $t^\ast_{pre}$ and ${t'}^\ast_{pre}$.
At those times, the active commands are both low and, hence, due to code agreement low the programs to be executed in both runs are the same, $\code (r,t^\ast_{pre}) = \code (r', {t'}^\ast_{pre})$. 

\begin{reasonbycases}
\item The active command in $\code (r,t^\ast_{pre})$, and hence in $ \code (r', {t'}^\ast_{pre})$ is a declassification assignment.\\
Consequently, the program in both runs has the form $\icontrol{x_ {src}}{x_{dest}};\cm{restp}$ and in the next execution step is reduced to $\cm{restp}$ according to Table~\ref{Tab:Runs} so that the longest low level suffixes are both  $\mathit{Lcont}(\cm{restp})$.
By choice of $t^\ast_{pre}$ and ${t'}^\ast_{pre}$, the low level execution steps at those times are observed, but afterwards there is no
observation of the high execution until $t$ and $t'$, respectively, according to Table~\ref{Tab:RunsObs}, so that it holds $\obsst (r , t^\ast_{pre} +1) = \obsst (r , t)$ and $\obsst (r' , {t'}^\ast_{pre} + 1) = \obsst (r',t')$, respectively.
Due to the precondition $\obsst (r ,t ) = \obsst (r' , t')$ the same value is written to low memory in $(r,{t}^\ast_{pre})$ and $(r',{t'}^\ast_{pre})$
and thus state agreement for $\Q{t^\ast_{pre}+1}{{t'}^\ast_{pre}+1}$ is fulfilled by induction hypothesis.
\item 
The active command in $\code (r,t^\ast_{pre})$, and hence in $ \code (r', {t'}^\ast_{pre})$ is not a declassification assignment.\\
By Property~\ref{Def:NoReadUp} (No-read-up), from state agreement low in $(r,t^\ast_{pre})$ and $(r',{t'}^\ast_{pre})$ we obtain that 
in the next step of both runs,  at times $t^\ast_{pre}+1$ and ${t'}^\ast_{pre}+1$, respectively,  the program codes are still equal
 and so are the longest low level suffixes, that is $\mathit{Lcont}( \code (r, t_{pre}^\ast + 1) )= \mathit{Lcont}( \code (r', {t'}_{pre}^\ast +1))$.
 
 Moreover, the latter argument with Property~\ref{Def:NoReadUp} implies that the low memory states in both runs still agree and hence state agreement is fulfilled. 
 \end{reasonbycases}
 
 Monotonicity and completeness hold by construction of the extended correspondence.
 With the previous arguments, we have verified all properties of a correspondence relation for the extension.
\item $t_{pre}^\ast +1 = t$  and ${t'}^\ast_{pre} + 1 = t'$.\\
Again we extend the correspondence to $\Q{t_{pre}^\ast +1}{{t'}^\ast_{pre} + 1}$.
By the precondition of the lemma, the  active commands in  $(r,t_{pre}^\ast +1)$ and $(r',{t'}_{pre}^\ast +1)$ are low and we can proceed as in the previous case to show $\code (r,t_{pre}^\ast +1) =  \code (r',{t'}_{pre}^\ast +1)$ (code agreement low) and the other properties of correspondence relations.
\item $t_{pre}^\ast +1 = t$  and ${t'}^\ast_{pre} + 1 <  t'$, or $t_{pre}^\ast +1 < t$  and ${t'}^\ast_{pre} + 1 = t'$.\\
These cases cannot occur because as we have just argued the programs in the next steps of both runs,  at $t^\ast_{pre}+1$ and ${t'}^\ast_{pre}+1$, respectively, are equal, 
but before $t$ and $t'$, respectively, the active command is high and at $t$ and $t'$, respectively, it is low. 
\end{reasonbycases}

Next, we use an inductive argument for each $t_{pre} > t_{pre}^\ast +1$  and $t'_{pre}> {t'}_{pre}^\ast +1$ until we reach $t$ and $t'$ respectively.
\begin{reasonbycases}
\item\label{Case:Ind-CorrespondEx} $t_{pre}\neq t$ and $t'_{pre} \neq t'$.\\
By choice of $t^\ast_{pre}$ and ${t'}^\ast_{pre}$, the active commands in $(r,t_{pre})$ and $(r',t'_{pre})$ are high.
We extend the correspondence from $r$ to $r'$ until $t_{pre}-1$ and $t'_{pre}-1$ to $\Q{t_{pre}}{t'_{pre}}$ and justify that for this extension the properties of a correspondence relation are still fulfilled. Level agreement holds by construction.

By choice of $t_{pre}$ and $t'_{pre}$,  in the previous steps, $t_{pre}-1 > t_{pre}^\ast$ and $t'_{pre}-1 > {t'}^\ast_{pre}$, the active commands are high.
Therefore, Property~\ref{Def:NoWriteDown} (No-write-down) ensures that the transition from the previous step to $t_{pre}$ and $t'_{pre}$, respectively, does not change low memory. Hence, state agreement is still fulfilled.

Code agreement high holds at the previous steps and hence also at $t_{pre}$ and $t'_{pre}$. Finally, monotonicity and completeness hold by construction
of the extended correspondence.
\item $t_{pre}\neq t$ and $t'_{pre} = t'$.\\
By choice of $t^\ast_{pre}$, the active command in $(r,t_{pre})$  is high, whereas by the precondition of the lemma in $(r',t'_{pre})$ it is low.
We extend the correspondence from $r$ to $r'$ until $t_{pre}-1$ and $t'_{pre}-1$ to $\Q{t_{pre}}{t'_{pre}-1}$ and justify that for this extension the properties of a correspondence relation are still fulfilled. 

In the previous step in $r'$, $t'_{pre}-1 > {t'}^\ast_{pre}$, the active command is high so that level agreement is still fulfilled for the extension.
Also the active command in $(r,t_{pre}-1)$ is high.
Therefore,  the transition from the previous step to $t_{pre}$ in $r$ does not change low memory by Property~\ref{Def:NoWriteDown} (No-write-down).
Moreover, since by the inductive argument there is a correspondence $\Q{t_{pre}-1}{t'_{pre}-1}$, the low memory state is the same $\mem_\low (r, t_{pre}-1) = \mem_\low (r', t'_{pre}-1)$ and hence also $\mem_\low (r, t_{pre}) = \mem_\low (r', t'_{pre}-1)$.
This way, we have just argued that state agreement is fulfilled for the extended correspondence.

Finally,  code agreement high holds for $\Q{t_{pre}-1}{t'_{pre}-1}$ and hence still for $\Q{t_{pre}}{t'_{pre}-1}$. 
Monotonicity and completeness are fulfilled by construction.
\item $t_{pre} = t$ and $t'_{pre} \neq t'$.\\
This case can be treated analogously to the previous one.
\item $t_{pre} = t$ and $t'_{pre} = t'$.\\
In this case,  the active commands in both runs are low by precondition of the lemma. 
So we extend the correspondence to $\Q {t}{t'}$ and show the required properties. 
Level agreement holds by construction.

Since for our inductive argument we consider $t-1 > t^\ast_{pre}$ and $t'-1 > {t'}^\ast_{pre}$, the active command in the previous step 
$t-1$ and $t'-1$, respectively,  is high so that we can proceed as in Case~\ref{Case:Ind-CorrespondEx}.
\end{reasonbycases}
\end{appendix}
\end{document}